\documentclass[10pt,british,tightenlines,eqsecnum,floats,aps,amsmath,amssymb,nofootinbib,superscriptaddress,prd,showpacs]{revtex4}
\usepackage[utf8]{inputenc}
\usepackage{babel}
\usepackage{amsmath}
\usepackage{amssymb}
\usepackage{graphicx}
\usepackage{esint}
\usepackage[unicode=true,pdfusetitle,
 bookmarks=true,bookmarksnumbered=false,bookmarksopen=false,
 breaklinks=false,pdfborder={0 0 1},backref=false,colorlinks=false]
 {hyperref}

\makeatletter

\providecommand{\tabularnewline}{\\}

\@ifundefined{textcolor}{}
{%
 \definecolor{BLACK}{gray}{0}
 \definecolor{WHITE}{gray}{1}
 \definecolor{RED}{rgb}{1,0,0}
 \definecolor{GREEN}{rgb}{0,1,0}
 \definecolor{BLUE}{rgb}{0,0,1}
 \definecolor{CYAN}{cmyk}{1,0,0,0}
 \definecolor{MAGENTA}{cmyk}{0,1,0,0}
 \definecolor{YELLOW}{cmyk}{0,0,1,0}
}

\@ifundefined{definecolor}
 {\usepackage{color}}{}

\usepackage{euscript}\usepackage{epsfig}

\usepackage{amsfonts}
\usepackage{fancyhdr}

\makeatother

\begin{document}

\title{Gravitational collapse with tachyon field and barotropic fluid}

\author{Y. Tavakoli}

\email{tavakoli@ubi.pt}

\selectlanguage{british}%

\affiliation{Departamento de Física, Universidade da Beira Interior, 6200 Covilhã,
Portugal}

\author{J. Marto}

\email{jmarto@ubi.pt}

\selectlanguage{british}%

\affiliation{Departamento de Física, Universidade da Beira Interior, 6200 Covilhã,
Portugal}

\author{A. H. Ziaie}

\email{ahziaie@sbu.ac.ir}

\selectlanguage{british}%

\affiliation{Department of Physics, Shahid Beheshti University, Evin, 19839 Tehran,
Iran}

\author{P. V. Moniz}

\email{pmoniz@ubi.pt}

\selectlanguage{british}%

\affiliation{Departamento de Física, Universidade da Beira Interior, 6200 Covilhã,
Portugal}

\affiliation{CENTRA, IST, Av. Rovisco Pais, 1, Lisboa, Portugal}
\begin{abstract}
A particular class of space-time, with a tachyon field, $\phi$, and
a barotropic fluid constituting the matter content, is considered
herein as a model for gravitational collapse. For simplicity, the
tachyon potential is assumed to be of inverse square form \emph{i.e.},
$V(\phi)\sim\phi^{-2}$. Our purpose, by making use of the specific
kinematical features of the tachyon, which are rather different from
a standard scalar field, is to establish the several types of asymptotic
behavior that our matter content induces. Employing a dynamical system
analysis, complemented by a thorough numerical study, we find classical
solutions corresponding to a naked singularity or a black hole formation.
In particular, there is a subset where the fluid and tachyon participate
in an interesting tracking behaviour, depending sensitively on the
initial conditions for the energy densities of the tachyon field and
barotropic fluid. Two other classes of solutions are present, corresponding
respectively, to either a tachyon or a barotropic fluid regime. Which
of these emerges as dominant, will depend on the choice of the barotropic
parameter, $\gamma$. Furthermore, these collapsing scenarios both
have as final state the formation of a black hole. 
\end{abstract}


\date{\today}

\maketitle

\section{Introduction}

\label{intro} The study of the final state of the gravitational collapse
of initially regular distributions of matter is one of the open problems
in classical general relativity, having attracted remarkable attention
in past decades. When a sufficiently massive star exhausts all the
thermonuclear sources of its energy, it would undergo a collapsing
scenario due to its own gravity, without reaching a final state in
terms of a neutron star or white dwarf. Under a variety of circumstances,
singularities will inevitably emerge (geodesic incompleteness in space-time),
matter densities and space-time curvatures diverging. Albeit the singularity
theorems \cite{2,Hawking} state that there exist space-time singularities
in a generic gravitational collapse, they provide no information on
the nature of singularities: the problem of whether these regions
are hidden by a space-time event horizon or can actually be observed,
remains unsolved. The cosmic censorship conjecture (CCC), as hypothesized
by Penrose \cite{Penrose}, conveys that the singularities appearing
at the collapse final outcome must be hidden within an event horizon
and thus no distant observer could detect them. A black hole forms.
Although the CCC plays a crucial role in the physics of black holes,
there is yet no proof of it, due to the lack of adequate tools to
treat the global characteristics of the field equations. Nevertheless,
in the past 30 years many solutions to the field equations have been
discovered, which exhibit the occurrence of naked singularities, where
the matter content has included perfect and imperfect fluids \cite{PFluid,imperfectF},
scalar fields \cite{sc}, self-similar models \cite{SS} and null
strange quarks \cite{NSQ}. Basically, it is the geometry of trapped
surfaces that decides the visibility or otherwise of the space-time
singularity. In case the collapse terminates into a naked singularity,
the trapped surfaces do not emerge early enough, allowing (otherwise
hidden) regions to be visible to the distant observers.

The gravitational collapse of scalar fields is of relevance \cite{LIvRev1},
owing to the fact that they are able to mimic other types of behaviours,
depending on the choice of the potentials. Scalar field models have
been extensively examined for studying CCC in spherically symmetric
models \cite{sphsym}, non-spherically symmetric models \cite{nonsphsym}
and also for static cases \cite{static}. Their role in understanding
the machinery governing the causal structure of space-time was available
since the 90's, when the numerical solutions exhibiting naked singularities
were found numerically by Choptuik \cite{Choptuik} and analytically
by Christodoulou \cite{Chris}. There are in the literature a few
papers discussing gravitational collapse in the presence of a scalar
field joined by a fluid for the matter content\cite{Pereira:2008yq,Jankiewicz:2006fh}:
in summary, a black hole forms in these collapsing situations. However,
to our knowledge, a tachyon scalar field has not yet been considered
regarding whether a black hole or naked singularity forms, that is
to say, in the CCC context, together with a fluid. Tachyon fields
arise in the framework of string theory \cite{tachst} and have been
of recent use in cosmology \cite{inftach}. The action for the tachyon
field has a non-standard kinetic term \cite{supers}, enabling for
several effects whose dynamical consequences are different from those
of a standard scalar field \cite{RLazkoz}. Namely, other (anti-)friction
features that can alter the outcome of a collapsing scenario. This
constitutes a worthy motivation to investigate the scenario where
a tachyon field is added to a barotropic fluid, both constituting
the matter content present in the collapse process: on the one hand,
the fluid will play the role of conventional matter from which a collapse
can proceed into, whereas, on the other hand, the tachyon would convey,
albeit by means of a simple framework, some intrinsic features from
a string theory setting. Restricting ourselves herein to the tachyon
as the intrinsic string ingredient influencing the collapse, let us
nevertheless point that many other string features could be incorporated
in subsequent similar studies \cite{LIvRev1,tachst,supers}. Our purpose,
in this paper, by investigating the gravitational collapse of a barotropic
fluid together with a tachyon field, is therefore to establish the
types of final state that can occur (i.e., whether a black hole or
a naked singularity emerges, in the context of the CCC), which matter
component will determine the outcome. In particular, if the late time
tachyon behaviour, possibly competing with the fluid and eventually
becoming dominant, could allow interesting features to appear.

We then organize this paper as follows. In section \ref{collapse}
we give a brief review on the gravitational collapse of a specific
space-time, namely the marginally bounded case (cf. \cite{4}). In
section \ref{Classic} we study, by means of a dynamical system analysis,
the gravitational collapse employing a tachyon and a barotropic fluid
as the matter content. The analytical study is complemented by a careful
numerical investigation. In section \ref{discussion} we present our
conclusions and a discussion of our results.

\section{A briefing on marginally bound gravitational collapse}

\label{collapse}

In this section, we will discuss the space-time region inside the
collapsing sphere which will contain the chosen matter content. An
isotropic Friedmann-Robertson-Walker (FRW) metric, in comoving coordinates,
will be considered as the interior space-time for the gravitational
collapse. However, in order to study the whole space-time, we must
match this interior region to a suitable exterior. In the model herein,
it is convenient to consider a spherically symmetric and inhomogeneous
space-time such as the Schwarzschild or the generalized Vaidya geometries
to model the space-time outside the collapsing sphere.

The interior space-time for the marginally bounded case (cf. \cite{4})
can be parametrized as 
\begin{align}
ds^{2}=-dt^{2}+a(t)^{2}dr^{2}+R(t,r)^{2}d\Omega^{2},\label{metric}
\end{align}
 where $t$ is the proper time for a falling observer whose geodesic
trajectories are labeled by the comoving radial coordinate $r$, $R(t,r)=ra(t)$
is the area radius of the collapsing volume and $d\Omega^{2}$ being
the standard line element on the unit two-sphere. The Einstein's field
equations for this model can be presented \cite{2} as (we have set
$8\pi G=c=1$ throughout this paper) 
\begin{equation}
\rho=\frac{F_{,r}}{R^{2}R_{,r}},~~~~p=\frac{-\dot{F}}{R^{2}\dot{R}}\ ,\label{einstein-1}
\end{equation}
 where the mass function 
\begin{equation}
F(t,r)=R\dot{R}^{2},\label{einstein2}
\end{equation}
 is the total gravitational mass within the shell, labeled by the
comoving radial coordinate $r$ with ``$,r\equiv\partial_{r}"$ and
``$\cdot\equiv\partial_{t}$''. Since we are interested in a continuous
collapsing scenario, we take $\dot{a}<0$, implying that the area
radius of the collapsing shell, for a constant value of $r$, decreases
monotonically. Splitting the above metric for a two-sphere and a two
dimensional hypersurface, normal to the two-sphere, we have 
\begin{equation}
ds^{2}=h_{ab}dx^{a}dx^{b}+R(t,r)^{2}d\Omega^{2},~~~~h_{ab}={\rm diag}\left[-1,a(t)^{2}\right],\label{metric1}
\end{equation}
 whereby the Misner-Sharp energy \cite{MS} is defined to be 
\begin{align}
E(t,r)=\frac{R(t,r)}{2}\left[1-h^{ab}\partial_{a}R\partial_{b}R\right]=\frac{R\dot{R}^{2}}{2}\ .\label{Misner-Sharp}
\end{align}
 To discuss the final state of gravitational collapse it is important
to study the conditions under which the trapped surfaces form. From
the above definition, it is the ratio $2E(t,r)/R(t,r)$ that controls
the formation of trapped surfaces \cite{MS} (note that the mass function
defined above is nothing but twice the Misner-Sharp mass). Introducing
the null coordinates 
\begin{align}
d\xi^{+}=-\frac{1}{\sqrt{2}}\left[dt-a(t)dr\right],~~~~d\xi^{-}=-\frac{1}{\sqrt{2}}\left[dt+a(t)dr\right],\label{doublenull}
\end{align}
 the metric (\ref{metric1}) can be recast into the double-null form
\begin{align}
ds^{2}=-2d\xi^{+}d\xi^{-}+R(t,r)^{2}d\Omega^{2}.\label{metricdnull}
\end{align}
 The radial null geodesics are given by $ds^{2}=0$. We then may deduce
that there exists two kinds of null geodesics which correspond to
$\xi^{+}=constant$ and $\xi^{-}=constant$, the expansions of which
are given by 
\begin{align}
\Theta_{\pm}=\frac{2}{R}\partial_{\pm}R,\label{expansion}
\end{align}
 where 
\begin{align}
\partial_{+}=\frac{\partial}{\partial\xi^{+}}=-\sqrt{2}\left[\partial_{t}-\frac{\partial_{r}}{a(t)}\right],~~~\partial_{-}=\frac{\partial}{\partial\xi^{-}}=-\sqrt{2}\left[\partial_{t}+\frac{\partial_{r}}{a(t)}\right].\label{p+p_}
\end{align}
 The expansion parameter measures whether the bundle of null rays
normal to the sphere is diverging $(\Theta_{\pm}>0)$ or converging
$(\Theta_{\pm}<0)$. The space-time is referred to as trapped, untrapped
and marginally trapped if 
\begin{align}
\Theta_{+}\Theta_{-}>0,~~~~\Theta_{+}\Theta_{-}<0,~~~~\Theta_{+}\Theta_{-}=0,\label{sp}
\end{align}
 respectively, where we have noted that $h^{ab}\partial_{a}R\partial_{b}R=-\frac{R^{2}\Theta_{+}\Theta_{-}}{2}$.
The third case characterizes the outermost boundary of the trapped
region, the apparent horizon equation being $\dot{R}^{2}=1$. From
the point of view of equation (\ref{Misner-Sharp}), we see that at
the boundary of the trapped region, $2E(t,r)=R(t,r)$. Thus, in the
regions where the mass function satisfies the relation $F(t,r)>R(t,r)$,
the trapping of light does occur, whereas for the regions in which
the mass function is less than the area radius i.e., $F(t,r)<R(t,r)$,
no trapping happens \cite{MS}. In other words, in the former there
exists a congruence of future directed null geodesics emerging from
a past singularity and reaching to distant observers, causing the
singularity to be exposed (a naked singularity forms). In the latter,
no such families exist and trapped surfaces will form early enough
before the singularity formation: regions will be hidden behind the
event horizon, which lead to the formation of a black hole \cite{2}
(a similar interpretation has been given in \cite{tpsingh,malec}).

In order to further illustrate this specific gravitational collapse
process, let us employ a very particular class of solutions. We assume
that the behavior of the matter density, for $a\ll1$, is given by
the following ansatz%
\footnote{Notice that, since the interior space-time (\ref{metric1}) is spatially
homogeneous, it is obvious that, the energy-momentum tensor must also
be spatially homogeneous.%
} (cf. ref. \cite{goswamijoshi,HKY}), 
\begin{equation}
\rho(t)\approx\frac{c}{a^{n}},\label{energyapprox}
\end{equation}
 where $c$ and $n$ are positive constants%
\footnote{Using the energy density (\ref{energyapprox}) and pressure (\ref{pressapprox}),
the weak energy condition can be written as follows 
\begin{equation}
\rho+p=\frac{nc}{3a^{n}}>0.\label{WEC}
\end{equation}
 The weak energy condition is satisfied throughout the collapse process
(see also next section for more details regarding the issue of the
energy conditions).%
}. Integration of the first equation in (\ref{einstein-1}) gives the
following relation for the mass function as 
\begin{equation}
F=\frac{1}{3}\rho(t)R^{3}.\label{massfunction}
\end{equation}
 Hence we have 
\begin{equation}
\frac{F}{R}=\frac{c}{3}r^{2}a^{2-n}.\label{ratio}
\end{equation}

It is therefore possible to identify the following outcomes: 
\begin{itemize}
\item For $0<n<2$, the ratio $F/R<1$ is less than unity throughout the
collapse and if no trapped surfaces exist initially, due to the regularity
condition ($F(t_{0},r)/R(t_{0},r)<1$ for a suitable value of $c$
and $r$), then no trapped surfaces would form until the epoch $a(t_{s})=0$.
More precisely, there exists a family of radial null trajectories
emerging from a naked singularity. 
\item For $n\geq2$, the ratio $F/R$ goes to infinity and trapped surfaces
will form as the collapse evolves, which means that singularity will
be covered and no radial geodesics can emerge from it. Thus, a black
hole forms. 
\end{itemize}
From equation (\ref{energyapprox}), we may easily find a relation
between the energy density and pressure as 
\begin{equation}
p=\frac{n-3}{3}\rho,\label{pressapprox}
\end{equation}
 which shows that for $n<2$, where a naked singularity occurs, the
pressure gets negative values \cite{PSJRG}. From equations (\ref{einstein2}),
(\ref{energyapprox}) and bearing in mind the continuous collapse
process $(\dot{a}<0)$ we can solve for the scale factor as 
\begin{equation}
a(t)=\left(a_{0}^{\frac{n}{2}}+\frac{n}{2}\sqrt{\frac{c}{3}}(t_{0}-t)\right)^{\frac{2}{n}},\label{scalefactor}
\end{equation}
 where $t_{0}$ is the time at which the energy density begins increasing
as $a^{-n}$ and $a_{0}$ corresponds to $t_{0}$. Assuming that the
collapse process initiates at $t_{0}=0$, the time at which the scale
factor vanishes corresponds to $t_{s}=\frac{2}{n}\sqrt{\frac{3}{c}}a_{0}^{n/2}$,
implying that the collapse ends in a finite proper time.

\section{Gravitational collapse with tachyon and barotropic fluid}

\label{Classic}

\begin{figure*}[t]
\begin{minipage}[c]{1\textwidth}%
\begin{center}
\includegraphics[width=2.5in]{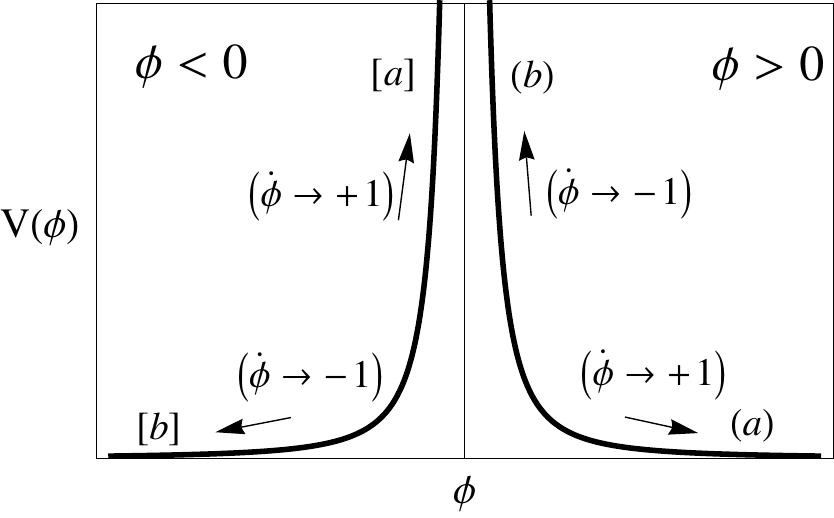} 
\par\end{center}%
\end{minipage}\caption{{\footnotesize The potential $V=V_{0}\phi^{-2}$, denoting the $\phi<0$
and $\phi>0$ branches as well as the asymptotic stages.}}

\label{pot-v2} 
\end{figure*}

The model we employ to study the gravitational collapse has an interior
space-time geometry as described above (see e.g. \cite{2,4,Joshi}).
Concerning the matter content, we consider a spherically symmetric
homogeneous tachyon field together with a barotropic fluid. We use
an inverse square potential for the tachyon field given by \cite{dbi,AGS,Frolov1},
\begin{equation}
V(\phi)=V_{0}\phi^{-2},\label{potential}
\end{equation}
 where $V_{0}$ is a constant. Note that we can consider the potential
(\ref{potential}) as two (mirror) branches (see figure \ref{pot-v2}),
upon the symmetry $\phi\rightarrow-\phi$, treating the $\phi>0$
and $\phi<0$ separately.

The total energy density of the collapsing system is therefore $\rho=\rho_{\phi}+\rho_{\text{b}}$,
with 
\begin{equation}
\rho=3H^{2}=\frac{V(\phi)}{\sqrt{1-\dot{\phi}^{2}}}+\rho_{\text{b}},\label{energy}
\end{equation}
 where $\rho_{\text{b}}$ is the energy density of the barotropic
matter, whose pressure $p_{\text{b}}$, in terms of the barotropic
parameter, $\gamma$, satisfies the relation $p_{\text{b}}=(\gamma-1)\rho_{\text{b}}$,
the barotropic parameter being positive, $\gamma>0$. Then, the Raychadhuri
equation for the collapsing system can be written as 
\begin{equation}
-2\dot{H}=\frac{V(\phi)\dot{\phi}^{2}}{\sqrt{1-\dot{\phi}^{2}}}+\gamma\rho_{\text{b}}.\label{Raychad1baro}
\end{equation}
 Furthermore, the equation of motion for the tachyon field can be
obtained as 
\begin{equation}
\ddot{\phi}=-(1-\dot{\phi}^{2})\left[3H\dot{\phi}+\frac{V_{,\phi}}{V}\right].\label{field}
\end{equation}
 The energy conservation of the barotropic matter is 
\begin{equation}
\dot{\rho}_{\text{b}}+3\gamma H\rho_{\text{b}}=0.\label{consenrbaro}
\end{equation}
 Integrating (\ref{consenrbaro}), the energy density for barotropic
matter can be presented as $\rho_{\text{b}}=\rho_{0\text{b}}a^{-3\gamma}$,
where $\rho_{0\text{b}}$ is a constant of integration. The total
pressure is given by 
\begin{equation}
p=p_{\phi}+(\gamma-1)\rho_{0\text{b}}a^{-3\gamma}.\label{press-a}
\end{equation}

In order to have a physically reasonable matter content for the collapsing
cloud, the tachyon field and the barotropic fluid would have to satisfy
the weak and dominant energy conditions%
\footnote{We note that satisfying the weak energy condition implies that the
null energy condition (NEC) is held as well.%
} \cite{2,Hawking,1}: for any time-like vector $V^{i}$, the energy-momentum
tensor $T_{ik}$ satisfies $T_{ik}V^{i}V^{k}\geq0$ and $T^{ik}V_{k}$
must be non-space-like. Since for the symmetric metric (\ref{metric})
we have $T_{t}^{t}=-\rho(t)$ and $T_{r}^{r}=T_{\theta}^{\theta}=T_{\varphi}^{\varphi}=p(t)$,
then, for the energy densities and the pressures of the tachyon field
and the barotropic fluid the energy conditions amount to the following
relations 
\begin{align}
 & \rho_{\phi}\geq0,\ \ \ \ \rho_{\phi}+p_{\phi}\geq0,\ \ \ \ |p_{\phi}|\leq\rho_{\phi},\label{EC1}\\
 & \rho_{\text{b}}\geq0,\ \ \ \ \rho_{\text{b}}+p_{\text{b}}\geq0,\ \ \ \ |p_{\text{b}}|\leq\rho_{\text{b}}.\label{EC2}
\end{align}
 The first two expressions in equations (\ref{EC1}) and (\ref{EC2})
denote the `weak energy condition' (WEC), and the last expression
denotes the `dominant energy condition' (DEC).

For a tachyon field, it is straightforward to show that the WEC is
satisfied: 
\begin{align}
\rho_{\phi}=\frac{V(\phi)}{\sqrt{1-\dot{\phi}^{2}}}\geq0,\ \ \ \ \text{and}\ \ \ \ \rho_{\phi}+p_{\phi}=\frac{V\dot{\phi}^{2}}{\sqrt{1-\dot{\phi}^{2}}}\geq0.\label{EC1a}
\end{align}
 In addition, since $\dot{\phi}^{2}<1$, then $V(\phi)(1-\dot{\phi}^{2})^{-1/2}\geq V(\phi)\sqrt{1-\dot{\phi}^{2}}$
from which it follows that $|p_{\phi}|\leq\rho_{\phi}$, and hence,
the DEC is satisfied for the tachyon matter.

For the barotropic fluid considered herein, we assume a regular initial
condition for the collapsing system, such as $\rho_{0b}>0$, indicating
that $\rho_{b}=\rho_{0b}a^{-3\gamma}>0$. In addition, since $p_{b}=(\gamma-1)\rho_{b}$,
then $p_{b}+\rho_{b}=\gamma\rho_{b}\geq0$ provided that $\gamma>0$;
through this paper we will consider the positive barotropic parameter
$\gamma$. Hence, the WEC is satisfied. On the other hand, considering
the DEC ($\rho_{b}\geq|p_{b}|$) it follows that the barotropic parameter
$\gamma$ must satisfy the range $\gamma\leq2$ (cf. the phase space
analysis in section \ref{Classic1}).

\subsection{Phase space analysis}

\label{Classic1}

We analyze in this (sub)section the gravitational collapse of the
above setup by employing a dynamical system perspective. We introduce
a new time variable $\tau$ (instead of the proper time $t$ present
in the comoving coordinate system $\{t,r,\theta,\varphi\}$), defined
as 
\begin{equation}
\tau\equiv-\log\left(\frac{a}{a_{0}}\right)^{3},\label{n-timeclass}
\end{equation}
 where $0<\tau<\infty$ (the limit $\tau\rightarrow0$ corresponds
to an initial condition of the collapsing system ($a\rightarrow a_{0}$)
and the limit $\tau\rightarrow\infty$ corresponds to $a\rightarrow0$).
For any time dependent function $f=f(t)$, we can therefore write%
\footnote{We recall that throughout this paper $H<0$, i.e., $\dot{a}<0$ is
assumed.%
} 
\begin{equation}
\frac{df}{d\tau}\equiv-\frac{\dot{f}}{3H},\label{dyn1}
\end{equation}
 We further introduce a new set of dynamical variables $x$, $y$
and $s$ as follows \cite{RLazkoz} 
\begin{equation}
x\equiv\dot{\phi},\qquad y\equiv\frac{V}{3H^{2}},\qquad s\equiv\frac{\rho_{\text{b}}}{3H^{2}}.\label{DyClXY}
\end{equation}

From the new variables (\ref{dyn1}), (\ref{DyClXY}) and the system
(\ref{Raychad1baro})-(\ref{consenrbaro}) with (\ref{potential}),
an autonomous system of equations is retrieved:
\begin{align}
\frac{dx}{d\tau} & \equiv f_{1}=\left(x-\frac{\lambda}{\sqrt{3}}\sqrt{y}\right)(1-x^{2}),\label{DyClXbaro}\\
\frac{dy}{d\tau} & \equiv f_{2}=-y\left[x\left(x-\frac{\lambda}{\sqrt{3}}\sqrt{y}\right)+s(\gamma-x^{2})\right],\label{DyClYbaro}\\
\frac{ds}{d\tau} & \equiv f_{3}=s(1-s)(\gamma-x^{2}).\label{DyClsbaro}
\end{align}
 Note that $\lambda$ is given by 
\begin{equation}
\lambda\equiv-\frac{V_{,\phi}}{V^{3/2}},\label{lambdanew}
\end{equation}
 which for the potential (\ref{potential}) brings $\lambda=\pm2/\sqrt{V_{0}}$
as a constant; for the $\phi>0$ branch, $\lambda>0$, and for $\phi<0$
branch, $\lambda<0$. Equation (\ref{energy}) in terms of the new
variables reads%
\footnote{Or equivalently, the surface where the trajectories will be present
can be written as $y^{2}=\left(1-s\right)^{2}\left(1-x^{2}\right)$.%
}, 
\begin{equation}
\frac{y}{\sqrt{1-x^{2}}}+s=1.\label{DyClFriedbaro}
\end{equation}
 The dynamical variables have the range $y>0,\: s\leq1$, and $-1<x<1$.
This brings an effective two-dimensional phase space%
\footnote{In the absence of a barotropic fluid, the effective phase space is
one-dimensional.%
}.

Setting $(f_{1},f_{2},f_{3})|_{(x_{c},y_{c},s_{c})}=0$, we can obtain
the critical points $(x_{c},y_{c},s_{c})$ for the autonomous system.
The stability can be subsequently discussed by using the eigenvalues
of the Jacobi matrix $\mathcal{A}$, defined at each fixed point $(x_{c},y_{c},s_{c})$,
as 
\begin{equation}
\mathcal{A}=\left(\begin{array}{ccc}
\frac{\partial f_{1}}{\partial x} & \frac{\partial f_{1}}{\partial y} & \frac{\partial f_{1}}{\partial s}\\
\frac{\partial f_{2}}{\partial x} & \frac{\partial f_{2}}{\partial y} & \frac{\partial f_{2}}{\partial s}\\
\frac{\partial f_{3}}{\partial x} & \frac{\partial f_{3}}{\partial y} & \frac{\partial f_{3}}{\partial s}
\end{array}\right)_{|(x_{c},y_{c},s_{c})}.\label{matrix2D}
\end{equation}
 Solutions, in terms of the dynamical variables, in the neighborhood
of a critical point $q_{i}^{\text{crit}}$ can be extracted by making
use of 
\begin{align}
q_{i}(t)\ =\ q_{i}^{\text{crit}}+\delta q_{i}(t),
\end{align}
 with the perturbation $\delta q_{i}$ given by 
\begin{align}
\delta q_{i}\ =\ \sum_{j}^{k}(q_{0})_{i}^{j}\exp(\zeta_{j}N),\label{pertQ1}
\end{align}
 where $q_{i}\equiv\{x,y,s\}$, $\zeta_{j}$ are the eigenvalues of
the Jacobi matrix, and the $(q_{0})_{i}^{j}$ are constants of integration.
We have summarized the fixed points for the autonomous system and
their more relevant stability properties in table \ref{T1}.

\begin{center}
\begin{table}[h!]
\begin{tabular}{|c|c|c|c|c|c|}
\hline 
~~point~~  & ~~~~$x$~~~~  & ~~~~$y$~~~~  & ~~~~$s$~~~~  & \qquad{}Existence\qquad{}  & ~~~~~~Stability~~~~~~ \tabularnewline
\hline 
$(a),\ [a]$  & $1$  & $0$  & $0$  & for all $\lambda$; $\gamma<1$  & Stable node\tabularnewline
 &  &  &  & for all $\lambda$; $\gamma>1$  & Saddle point\tabularnewline
$(b),\ [b]$  & $-1$  & $0$  & $0$  & for all $\lambda$; $\gamma<1$  & Stable node\tabularnewline
 &  &  &  & for all $\lambda$; $\gamma>1$  & Saddle point\tabularnewline
$(c),\ [c]$  & $-\frac{\lambda}{\sqrt{3}}\sqrt{y_{0}}$  & $y_{0}$  & $0$  & for all $\lambda$; $\gamma>\gamma_{1}$  & Unstable node\tabularnewline
 &  &  &  & for all $\lambda$; $\gamma\leq\gamma_{1}$  & Saddle point\tabularnewline
$(d),\ [d]$  & $0$  & $0$  & $1$  & for all $\lambda$; $\gamma\neq0$  & Saddle point \tabularnewline
 &  &  &  & for all $\lambda$; $\gamma=0$  & Unstable node\tabularnewline
$(e)$  & $-\sqrt{\gamma}$  & $\frac{3\gamma}{\lambda^{2}}$  & $s_{0}$  & {\footnotesize $\lambda>0$, $\gamma<\gamma_{1}<1$}  & Unstable node\tabularnewline
$[e]$  & $\sqrt{\gamma}$  & $\frac{3\gamma}{\lambda^{2}}$  & $s_{0}$  & {\footnotesize $\lambda<0$, $\gamma<\gamma_{1}<1$}  & Unstable node\tabularnewline
$\left(f\right),\ [f]$  & $1$  & $0$  & $1$  & for all $\lambda$; $\gamma>1$  & Stable node\tabularnewline
 &  &  &  & for all $\lambda$; $\gamma<1$  & Saddle point\tabularnewline
$\left(g\right),\ [g]$  & $-1$  & $0$  & $1$  & for all $\lambda$; $\gamma>1$  & Stable node\tabularnewline
 &  &  &  & for all $\lambda$; $\gamma<1$  & Saddle point\tabularnewline
$(h),\;[h]$ & $1$ & $0$ & $s_{1}$ & $\gamma=1$  & Stable node\tabularnewline
$(i),\;[i]$ & $-1$ & $0$ & $s_{1}$ & $\gamma=1$  & Stable node\tabularnewline
\hline 
\end{tabular}\caption{{\footnotesize Summary of critical points and their properties for
the $\phi<0$ (denoted with square brackets, i.e., as, for example,
$\left[a\right]$) and $\phi>0$ branches (denoted with regular brackets,
i.e., as, for example, $\left(a\right)$).} }

\label{T1} 
\end{table}

\par\end{center}

Table \ref{T1} is complemented with figure \ref{3D-stream}. The
autonomous system equations (\ref{DyClXbaro})-(\ref{DyClsbaro})
unfolds a three dimensional phase space constrained by (\ref{DyClFriedbaro}).
Therefore the main fixed points (and the trajectories nearby) are
located on surfaces of this three dimensional phase space. Furthermore,
from table \ref{T1}, we can focus our attention on three situations
characterized by having $s=0$, $s=1$ and $s=s_{0}$. Briefly, in
more detail: 
\begin{itemize}
\item \textbf{Point $(a)$}: The eigenvalues are $\zeta_{1}=-2$, $\zeta_{2}=-1$
and $\zeta_{3}=(\gamma-1)$. For $\gamma<1$, all the characteristic
values are real and negative, then the trajectories in the neighborhood
of this point are attracted towards it. Hence, $(a)$ is a stable
node (attractor). Finally, for $\gamma>1$, all characteristic values
are real, but one is positive and two are negative, the trajectories
approach this point on a surface and diverge along a curve: this is
a saddle point. 
\begin{figure*}[h]
\begin{centering}
\includegraphics[width=2in]{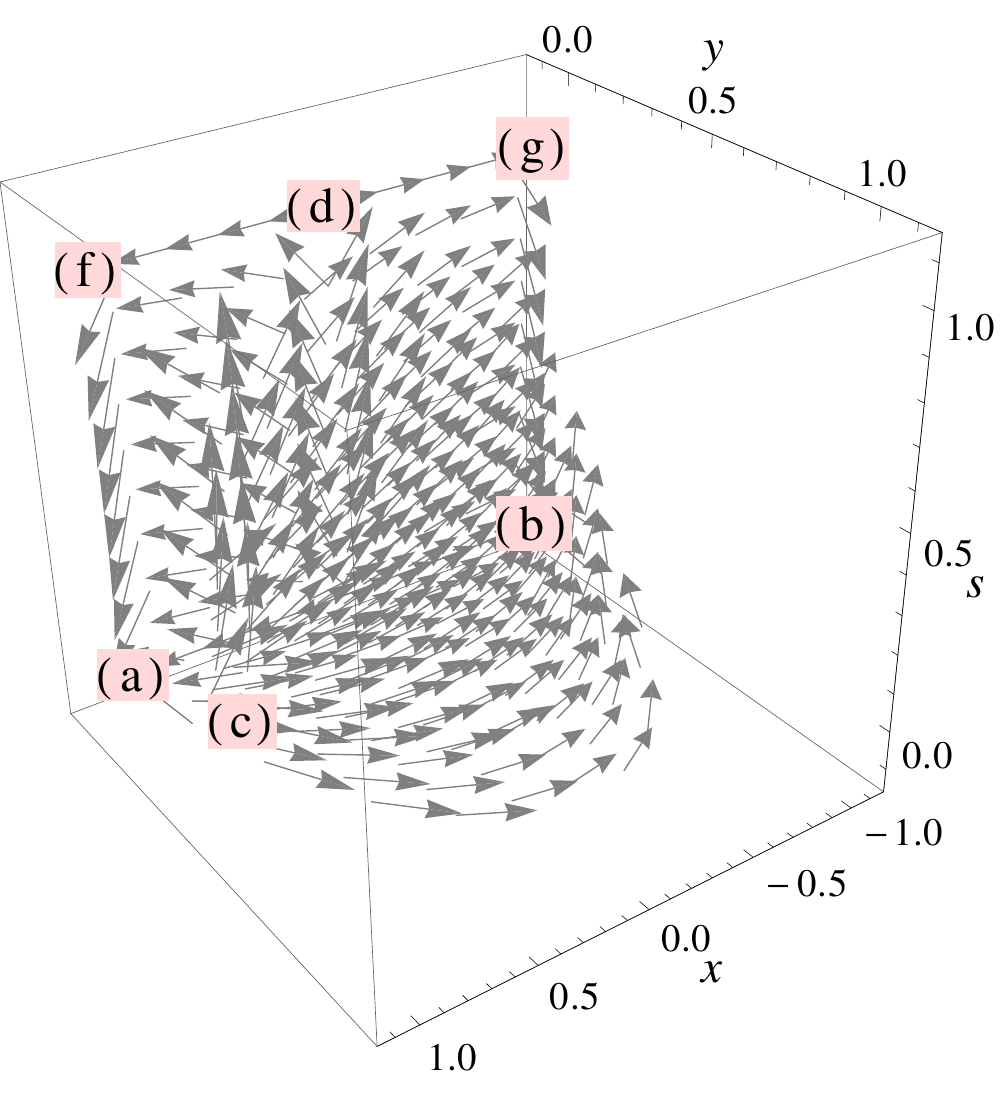}\quad{}\includegraphics[width=2in]{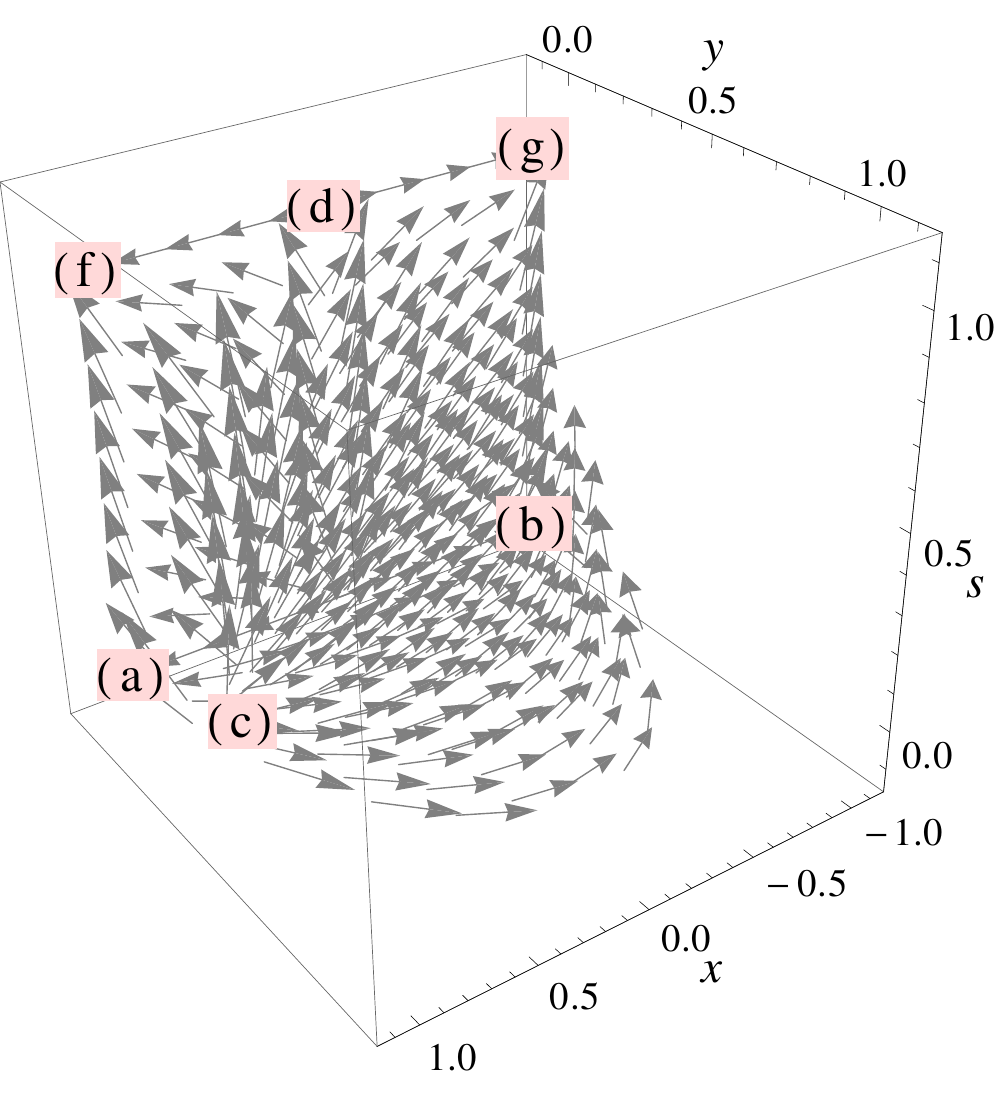} 
\par\end{centering}

\caption{{\footnotesize Trajectories in phase space and critical points: (i)
Left plot represents the phase space region $\left(x,y,s\right)$
constrained by (\ref{DyClFriedbaro}). Therein we also depicted all
the fixed points (see table \ref{T1}), except point $\left(e\right)$,
for $V_{0}=4/9$, $\gamma=0.5$. (ii) In the right plot we considered
the conditions as $V_{0}=4/9$, $\gamma=1.5$. We can illustrate that
going from $\gamma<1$ to $\gamma>1$ (from left to the right plot)
reverses the direction of the trajectories, i.e., in the left plot
the vector field is directed towards points $\left(a\right)$ or $\left(b\right)$.
In the right plot the vector field is directed towards points$\left(f\right)$
or $\left(g\right)$.}}

\label{3D-stream} 
\end{figure*}

\item \textbf{Point $(b)$}: The eigenvalues are $\zeta_{1}=-2$, $\zeta_{2}=-1$
and $\zeta_{3}=(\gamma-1)$. This point has the same eigenvalues of
point $(a)$ and similar asymptotic behavior, being also a stable
node for $\gamma<1$ (see the left plot in figure \ref{2D-stream})
and a saddle point for $\gamma>1$. 
\item \textbf{Point $(c)$:} This fixed point has eigenvalues $\zeta_{1}=0$,
$\zeta_{2}=(\lambda^{2}y_{0}+6y_{0}^{2})/6>0$ and $\zeta_{3}=(\gamma-\lambda^{2}y_{0}/3)$,
which all are real and $y_{0}=-\lambda^{2}/6+\sqrt{1+(\lambda^{2}/6)^{2}}$.
For $\gamma>\gamma_{1}\equiv\lambda^{2}y_{0}/3$, two components are
positive. However, from a numerical investigation we can assert that
this corresponds to an unstable saddle (see left plot in figure \ref{2D-stream}).
On the other hand, for $\gamma<\gamma_{1}$, one component is negative
and other is positive, and hence, a saddle point configuration would
emerge. 
\item \textbf{Point $(d)$:} The eigenvalues are $\zeta_{1}=1$, $\zeta_{2}=-\gamma$
and $\zeta_{3}=-\gamma$. As $\gamma>0$, trajectories approach this
point on a surface (the in-set) and diverge along a curve (the out-set).
This is a saddle\emph{ }point (see left plot in figure \ref{3D-stream}). 
\item \textbf{Point $(e)$:} This point is located at $(-\sqrt{\gamma},\frac{3\gamma}{\lambda^{2}},s_{0})$,
where $s_{0}=(1-\frac{3\gamma}{\lambda^{2}\sqrt{1-\gamma}})$. From
the constraint $0\leq s_{0}\leq1$ we get 
\begin{equation}
0\leq1-\frac{3\gamma}{\lambda^{2}\sqrt{1-\gamma}}\leq1,\label{const-track}
\end{equation}
 i.e., 
\begin{equation}
0\leq\gamma\leq\gamma_{1}<1\ .\label{const-track4}
\end{equation}
 When $\gamma\rightarrow0$, points $\left(e\right)$ and $\left(d\right)$
become coincident (see table \ref{T1}). When we consider $s_{0}=0$,
we obtain $\gamma=\gamma_{1}$, $x=-\frac{\lambda}{\sqrt{3}}\sqrt{y_{0}}$
and $y=y_{0}$. Consequently point $\left(e\right)$ and $\left(c\right)$
become coincident. Therefore, point $\left(e\right)$ can be found
along a curve that joins points $\left(c\right)$ (on the surface
with $s=0$) and $\left(d\right)$ (on the surface with $s=1$). The
eigenvalues are $\zeta_{1}=0$, with $\zeta_{2}$ and $\zeta_{3}$
given by, 
\begin{align}
\begin{aligned}\zeta_{2,3}= & \frac{1}{4}\left(2-\gamma\pm\sqrt{\left(1-\gamma\right)\left(4-16s_{0}\gamma\right)+\gamma^{2}}\right)\end{aligned}
.\label{eigencle}
\end{align}
 The eigenvalues (\ref{eigencle}) are non negative for $\gamma<\gamma_{1}$
and for $\gamma=\gamma_{1}$, $\zeta_{2}>0$ and $\zeta_{3}=0$.\medskip{}

\begin{figure*}[h]
\begin{minipage}[c]{1\textwidth}%
\begin{center}
\includegraphics[width=2.2in]{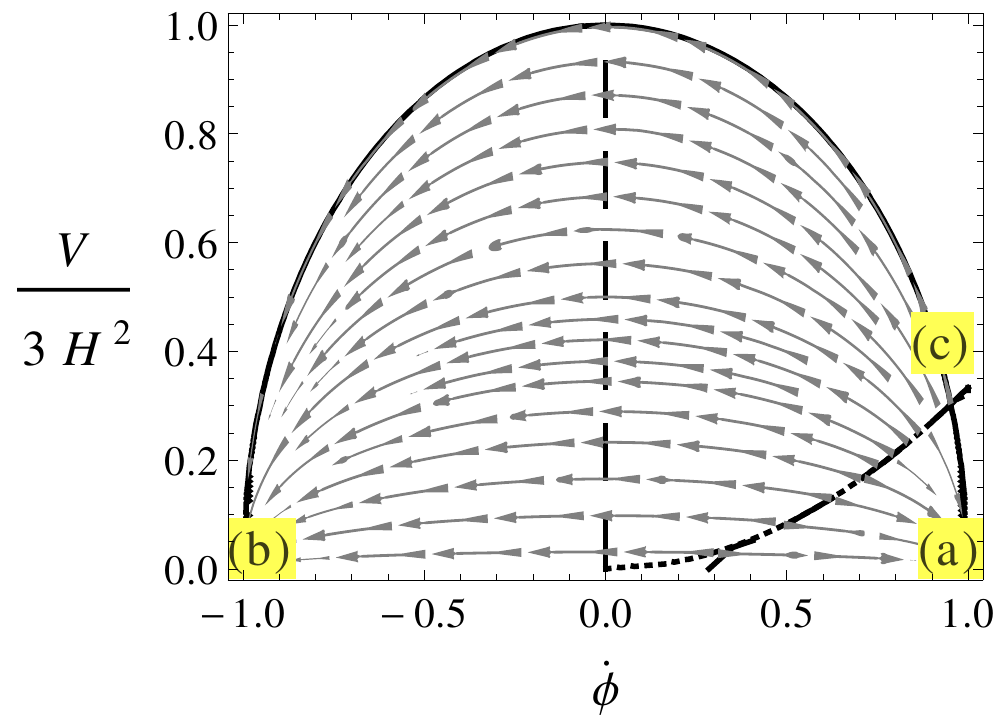}\quad{}\includegraphics[width=2.2in]{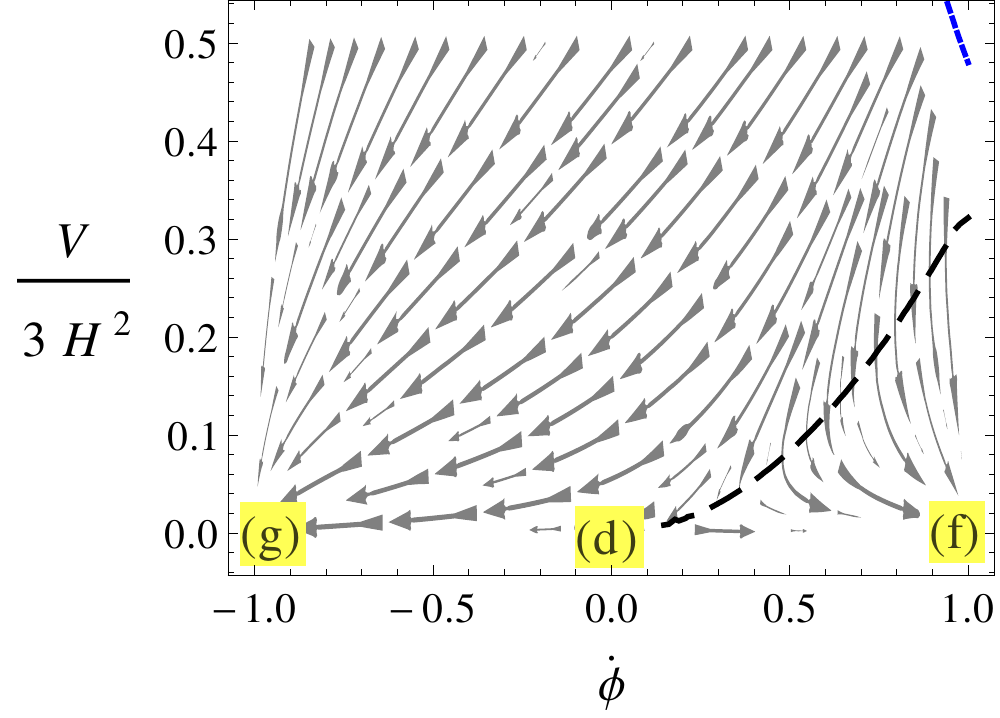}
\includegraphics[width=2.2in]{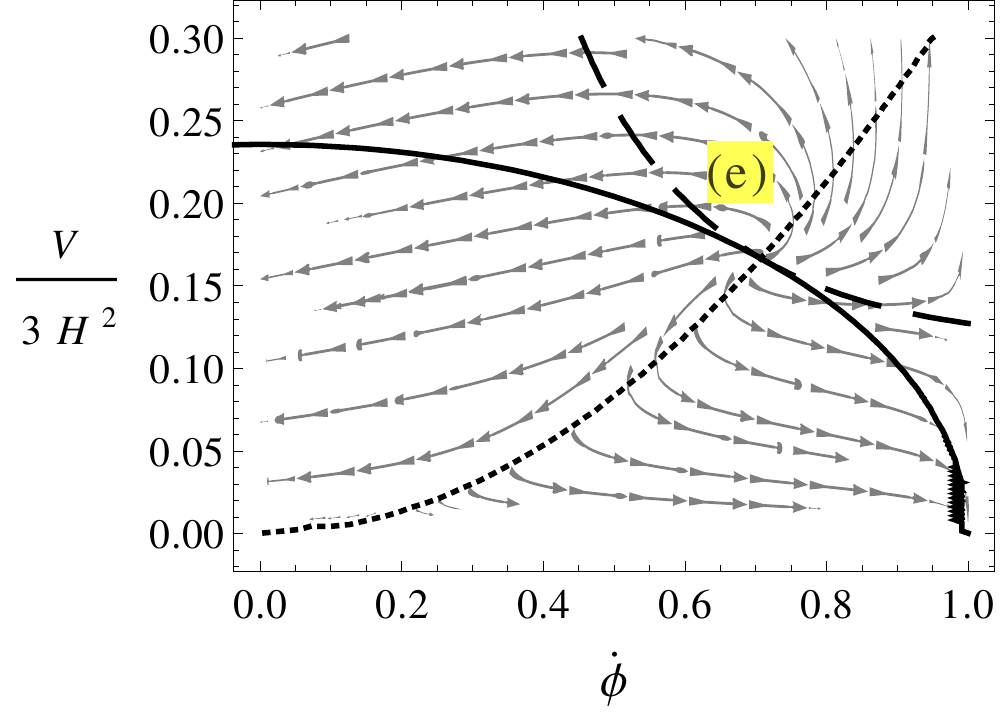} 
\par\end{center}%
\end{minipage}\caption{{\footnotesize (i) The top left plot represents a section, of the
three dimensional phase space presented in figure \ref{3D-stream},
labeled with $s=0$. The dashed and dotted lines are the zeros of
eqs. (\ref{DyClXbaro})-(\ref{DyClsbaro}). The full line represents
constraint (\ref{DyClFriedbaro}). Therefore, the fixed points are
found on their intersections. We considered the conditions as $V_{0}=4/9$,
$\gamma=0.5$. In this example we can locate the position of fixed
points $\left(a\right)$, $\left(b\right)$ and $\left(c\right)$
according to table \ref{T1}. (ii) The center plot represents a section,
showing fixed point $\left(e\right)$, labeled with $s=0.76$. The
dashed (and also dotted) lines are the zeros of eqs. (\ref{DyClXbaro})-(\ref{DyClsbaro})
and the full line represents constraint (\ref{DyClFriedbaro}), as
before. We considered the conditions as $V_{0}=4/9$, $\gamma=0.5$.
(iii) Finally, in the top right plot we represent a section, labeled
with $s=1$. We can identify fixed points $\left(g\right)$, $\left(d\right)$
and $\left(f\right)$. We considered the conditions as $V_{0}=4/9$,
$\gamma=1.2$.}}

\label{2D-stream} 
\end{figure*}

\medskip{}
 In the plot at the center of figure \ref{2D-stream} we provide a
section of the phase space showing point $\left(e\right)$, where
it can be seen that all trajectories are divergent from it. This behavior
is characteristic of an unstable node and expected whenever point
$\left(e\right)$ is found along the curve that connects $\left(c\right)$
and $\left(d\right)$.

\item \textbf{Point $(f)$:} The eigenvalues are $\zeta_{1}=-2$, $\zeta_{2}=-\gamma$
and $\zeta_{3}=(1-\gamma)$. For $\gamma>1$, all the characteristic
values are real and negative, then the trajectories in the neighborhood
of this point are attracted towards it. Hence, $(f)$ is a stable
node (attractor). Finally, for $\gamma<1$, all characteristic values
are real, but one is positive and two are negative. The trajectories
approach this point on a surface and diverge along a curve; this is
a saddle point. 
\item \textbf{Point $(g)$: }The eigenvalues are $\zeta_{1}=-2$, $\zeta_{2}=-\gamma$
and $\zeta_{3}=(1-\gamma)$. This point has the same eigenvalues of
point $(f)$ and similar asymptotic behavior, being also a stable
node for $\gamma>1$ (see the right plot in figure \ref{2D-stream})
and a saddle point for $\gamma<1$. 
\item \textbf{Point $(h)$:} This point is found along the line segment
with $x=1$, $y=0$ and $s\in\left]0,1\right[$. The eigenvalues are
$\zeta_{1}=-2$, $\zeta_{2}=-1$ and $\zeta_{3}=2(s_{1}-1)$, where
$s_{1}\in\left]0,1\right[$. In this case, all the characteristic
values are real and negative, then the trajectories in the neighborhood
of this points are attracted towards it. Hence, $(h)$ is a stable
node (attractor).
\item \textbf{Point $(i)$: }This point is found along the line segment
with $x=-1$, $y=0$ and $s\in\left]0,1\right[$. The eigenvalues
are $\zeta_{1}=-2$, $\zeta_{2}=-1$ and $\zeta_{3}=2(s_{1}-1)$.
This point has the same eigenvalues of point $(h)$ and similar asymptotic
behavior, being also a stable node. 
\end{itemize}
Before proceeding, let us add two comments. On the one hand, note
the transition that occurs, when going from $\gamma<1$ to $\gamma>1$,
an intermediate state as discussed (by means of a numerical study)
in subsection \ref{Tracking-solutions}. This bifurcation behaviour
is made explicit through the numerical methods employed. These allowed
us to confirm the results on the dynamical system analysis and to
further assess in regions like the transition from tachyon dominance
to fluid dominance, verifying all possible scenarios with those two
tools. On the other hand, figure \ref{3D-stream} deserves some attention
when we consider the trajectories approaching the stable nodes $(a)-(b)$
(left plot) and $(f)-(g)$ (right plot). Therein, the trajectories
approach $x\rightarrow\pm1$ along a segment line containing the stable
nodes. This situation implies that the energy density (\ref{energy})
diverges after reaching a point where $x\rightarrow\pm1$, $y\rightarrow0$
and $s\rightarrow s_{1}$. This observation will be important in the
next sections concerning the possible outcomes of the gravitational
collapse.

\subsection{Analytical and numerical results}

\label{Classic1-0}

Let us herewith discuss this section possible outcomes regarding the
collapsing system, employing elements from both our analytical as
well of numerical study.

\subsubsection{Tachyon dominated solutions}

\label{Tachyon dominated}

From the trajectories in the vicinity of $(a)$ and $(b)$, attractor
solutions can be described.

\begin{figure*}[h]
\begin{minipage}[c]{1\textwidth}%
\centering \includegraphics[width=2.1in]{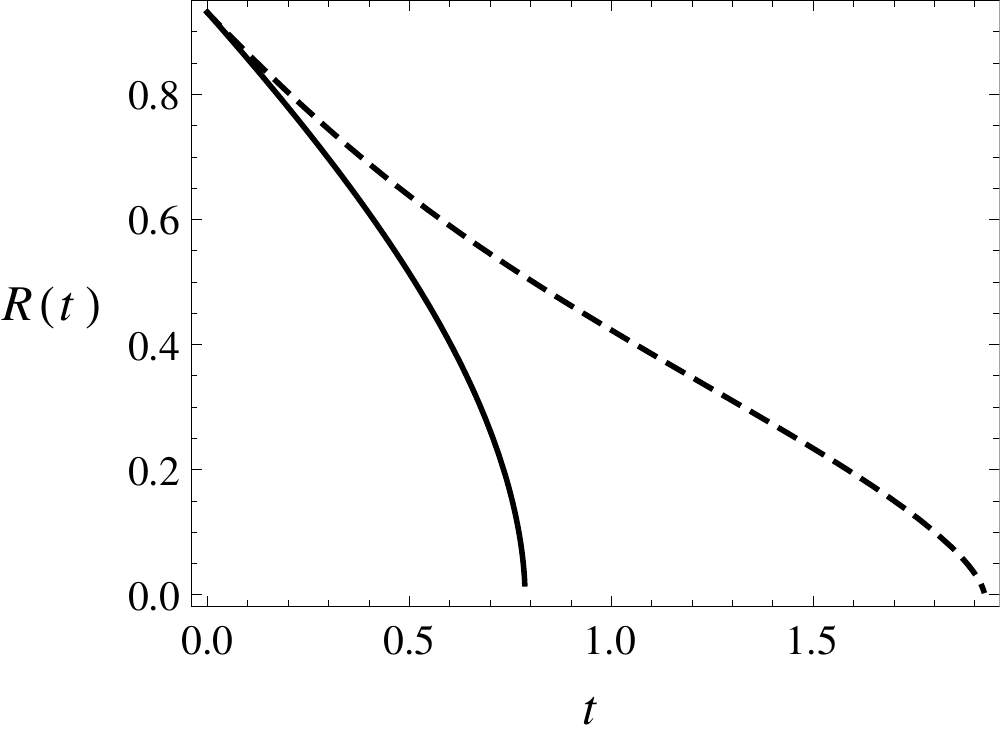}\quad{}\includegraphics[width=2.1in]{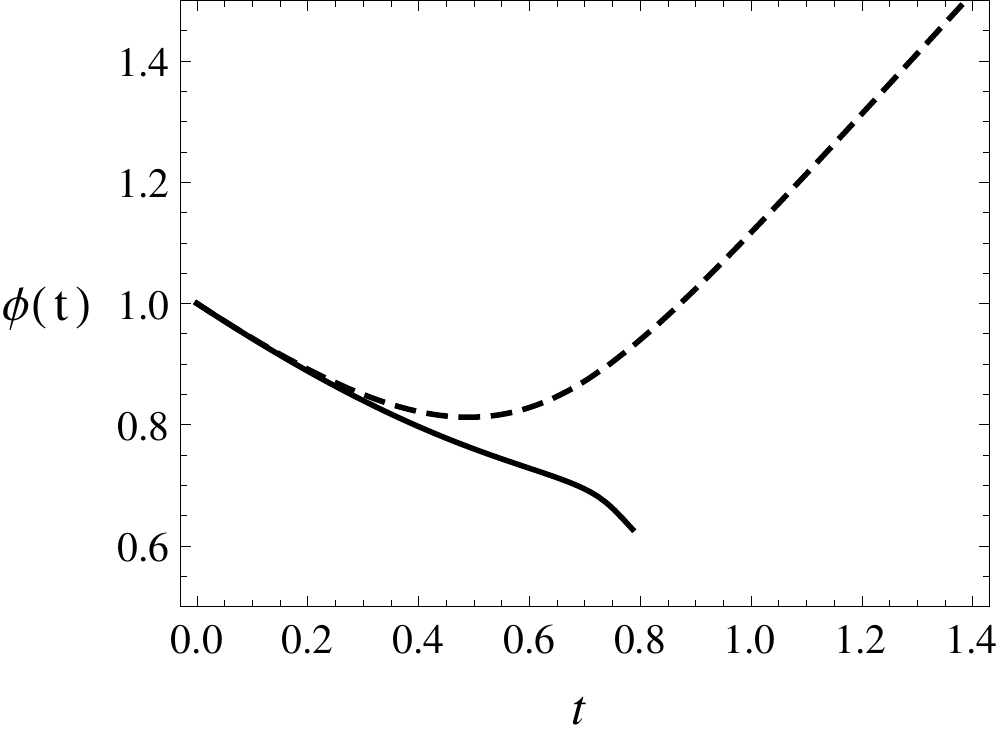}$\quad$\includegraphics[width=2.1in]{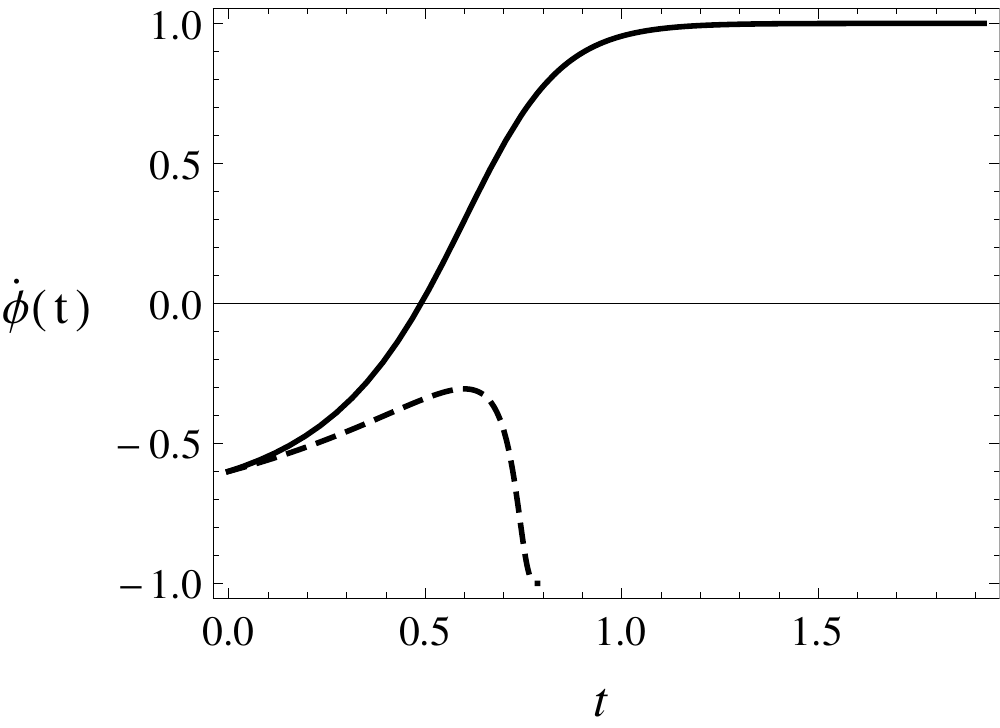} %
\end{minipage}\caption{{\footnotesize Behavior of the area radius, tachyon field and its
time derivative over time. We considered the initial conditions as:
$t_{i}=0$, $\rho_{0b}=1$, $V_{0}=4/9$, $a(0)=a_{0}$, $\dot{\phi}\left(0\right)=-0.6$
and $\phi_{0}=1$. We also have $\gamma=0.1$ (full lines) and $\gamma=1.2$
(dashed lines).}}

\label{F3b} 
\end{figure*}

The asymptotic behavior of $s$ near the point $(a)$ can be approximated
as $s\approx s_{c}+\exp(-\tau)=\exp(-\tau)$; hence, as $\tau\rightarrow\infty$
(i.e., $a\rightarrow0$), $s$ vanishes. Moreover, the time derivative
of the tachyon field is given by $\dot{\phi}\simeq1$, that is, the
tachyon field $\phi(t)$ has a linear time dependence and can be approximated
as (see figure \ref{F3b}) 
\begin{equation}
\phi(t)\ \simeq\ t+\phi_{0}.
\end{equation}
 \medskip{}

It should be noted that, for the $\phi>0$ branch, $\phi_{0}$ is
positive on the initial configuration of the collapsing system (where
$t=0$) and hence, the tachyon field increases with time, proceeding
downhill the potential. Within a finite amount of time, the tachyon
field reaches its maximum%
\footnote{The time at which the collapse reaches the singularity is finite.
Thus the tachyon field at the singularity remains finite as $\phi(t_{s})=t_{s}+\phi_{0}$
.%
} but finite value $\phi(t_{s})=\phi_{s}$ at $t_{s}<\phi_{0}$, with
the minimum (but non-zero value) $V\propto\phi_{s}^{-2}$. As the
tachyon potential decreases, the dynamical variable $y=\frac{V}{3H^{2}}$
vanishes.\medskip{}

\begin{figure*}[h]
\begin{minipage}[c]{1\textwidth}%
\centering \includegraphics[width=2in]{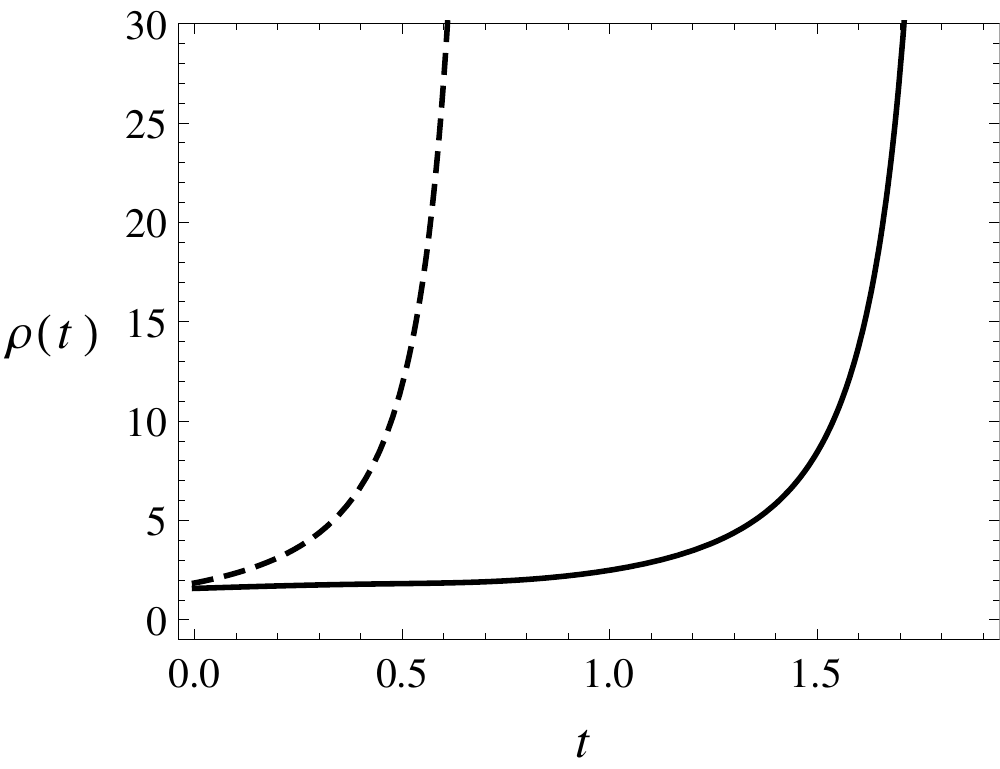}\quad{}\includegraphics[width=2in]{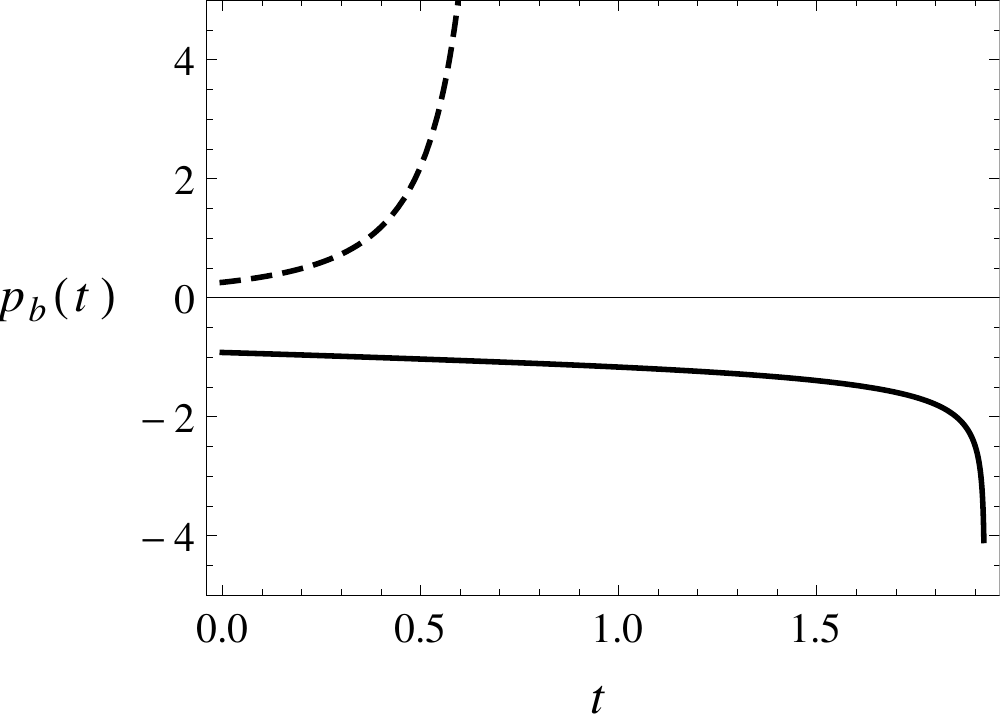}\quad{}\includegraphics[width=2in]{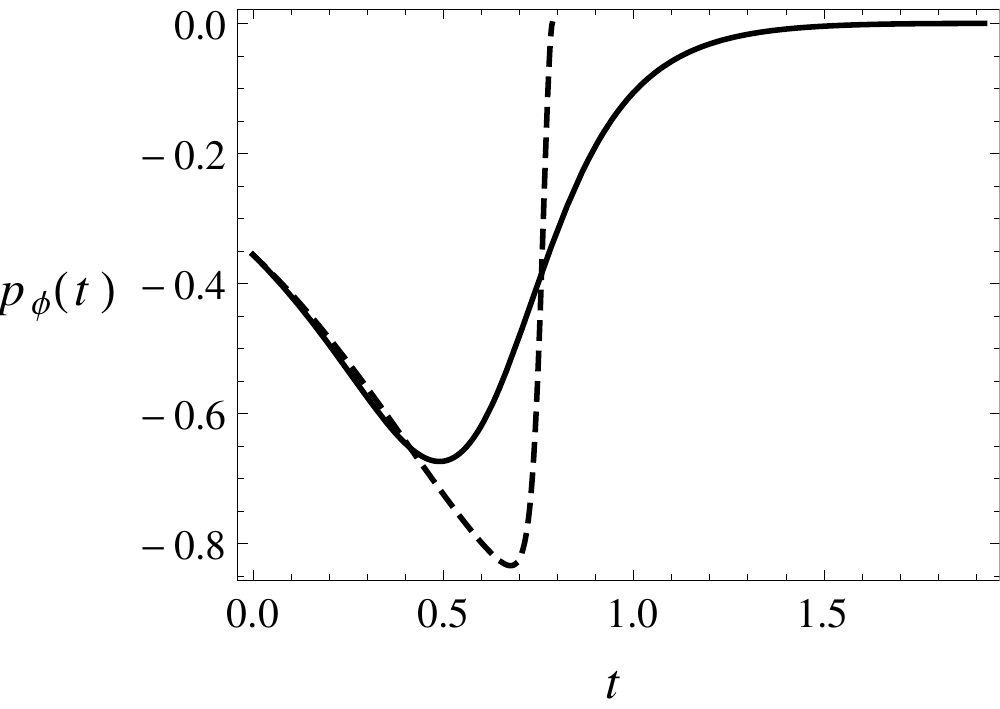} %
\end{minipage}\caption{{\footnotesize The energy density, barotropic pressure and tachyon
field pressure. We considered $t_{i}=0$, $a(0)=a_{0}$, $V_{0}=4/9$
and $\phi_{0}=0.6$ with $\gamma<1$ (full lines) and $\gamma>1$
(dashed lines). The total effective pressure ($p_{b}+p_{\phi}$) is
divergent and negative (for $\gamma<1$) in the final stage of the
collapse.}}

\label{F3} 
\end{figure*}

Thus, as $\dot{\phi}\rightarrow1$, the energy density of the system
diverges. Furthermore, the tachyon pressure $p_{\phi}=-V(\phi)(1-\dot{\phi}^{2})^{\frac{1}{2}}$
vanishes asymptotically%
\footnote{For the $\phi<0$ branch, $\phi_{0}$ is negative at the initial condition.
Thus, the absolute value of the tachyon field starts to decrease from
the initial configuration as $\phi(t)=t-|\phi_{0}|$ until the singular
point at time $t_{s}<|\phi_{0}|$, where tachyon field reaches its
minimum but non-zero value $\phi_{s}$. This leads it uphill the potential
until the singular epoch, where the potential becomes maximum but
finite.%
} (see figure \ref{F3} for plots of the energy density, barotropic
pressure and tachyon field pressure).\medskip{}

\begin{figure*}[h]
\begin{minipage}[c]{1\textwidth}%
\centering \includegraphics[width=2.1in]{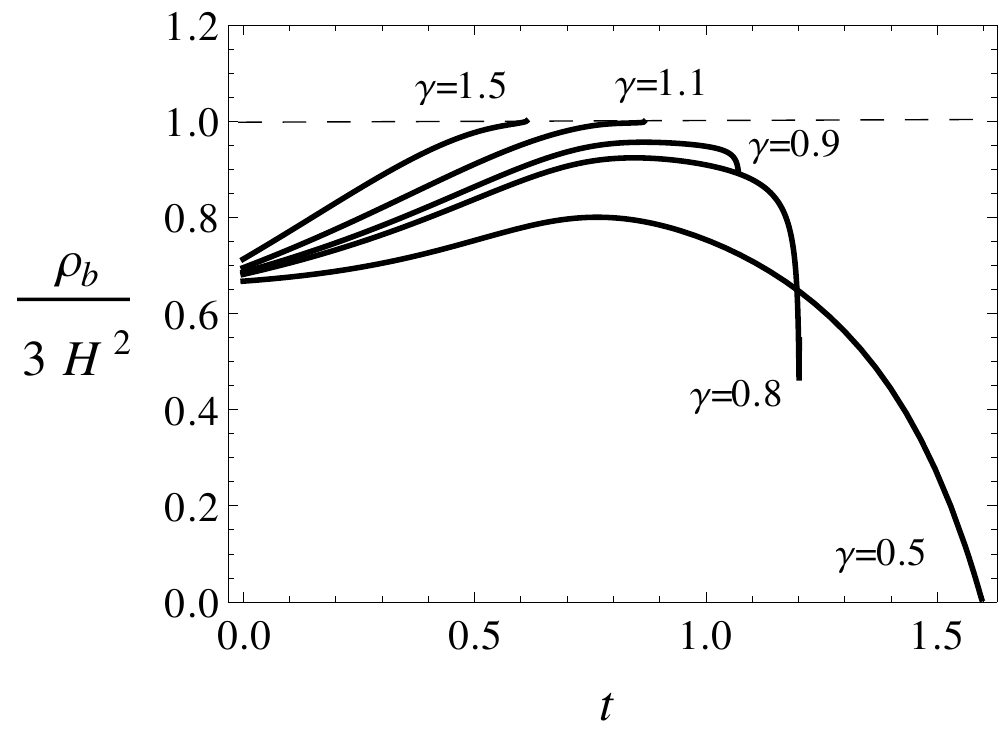}\quad{}\includegraphics[width=2.1in]{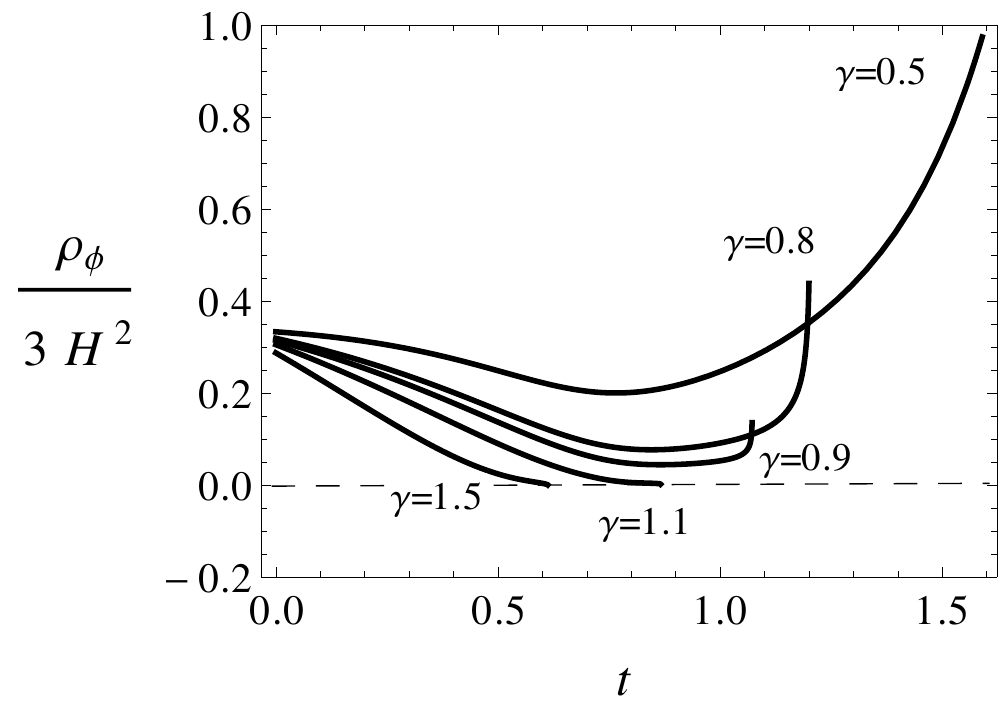} %
\end{minipage}\caption{{\footnotesize Behavior of the ratios $\frac{\rho_{b}}{3H^{2}}$ (left)
and $\frac{\rho_{\phi}}{3H^{2}}$ (right). We considered the initial
conditions as: $t_{i}=0$, $\rho_{0b}=1$, $V_{0}=4/9$, $a(0)=a_{0}$,
$\dot{\phi}\left(0\right)=-0.6$ and $\phi_{0}=1$. We illustrate
the transition between tachyon and fluid dominated solutions as a
function of the barotropic parameter. In these collapsing cases the
potential does not diverge and the tachyon field does not vanish.}}

\label{dominance} 
\end{figure*}

The total energy density of the collapsing system is given approximately
by the energy density of tachyon field (see figure \ref{dominance}
for numerical solutions), which corresponds to a dust-like matter
near the singularity as $\rho_{\phi}\propto1/a^{3}$ . From this relation
for the tachyon energy, we may write the energy density of the system
near point $(a)$, 
\begin{equation}
\rho\ \approx\ \rho_{o}a^{-3}.\label{Tenerg-a}
\end{equation}
 This induces a black hole at the collapse final state.

As far as $\left(b\right)$ is concerned, the time dependence of the
tachyon field can be obtained as 
\begin{equation}
\phi(t)\simeq-t+\phi_{0}.
\end{equation}
 For the $\phi>0$ branch, the tachyon field decreases from its initial
value at $t=0$ (where $\phi_{0}>0$), moving uphill the potential.
Then, the potential will reach a maximum (but finite) value when the
tachyon field reaches its minimum and nonzero value (i.e. $\phi\rightarrow\phi_{s}$
as $t\rightarrow t_{s}\leq\phi_{0}$). In the limit case $t_{s}=\phi_{0}$,
the tachyon field vanishes and the potential diverges. I.e., $y$
and $s$ vanish, the Hubble rate increases faster than the potential
and the barotropic fluid energy density diverges, that is, the total
energy density of the system asymptotically diverges.%
\footnote{On the other hand, for the $\phi<0$ branch, the tachyon field increases
from its initial condition as time evolves, proceeding to ever less
negative values as $\phi\rightarrow0^{-}$. In this case, it proceeds
downhill the tachyon potential till the system reaches $a=0$ at $t=t_{s}$,
where the Hubble rate and hence the total energy density of the system
diverge. This implies that the time at which the collapse system reaches
the singularity is always $t_{s}\leq\phi_{0}$.%
}. Likewise, when $\dot{\phi}\rightarrow1$ and $a\rightarrow0$, the
tachyon matter behaves as dust matter. The fate of the collapse for
this fixed point is as well a black hole formation.

The asymptotic solutions provided by the fixed points $(a)$ and $(b)$
correspond to a dust-like solution with a vanishing pressure for the
tachyon field, whose energy density reads $\rho_{\phi}\propto1/a^{3}$.
This is consistent with the WEC and DEC for the tachyon matter being
satisfied, as was mentioned before. Concerning the status of the energy
conditions for the barotropic fluid, as we indicated before, regularity
of the initial data for the collapsing matter respects the WEC. On
the other hand, stability of the solution in this case ensures that
$\gamma<1$, which satisfies the sufficient condition for the DEC.

\subsubsection{Fluid dominated solutions}

\label{Fluid dominated}

From the trajectories in the vicinity of $(f)$ and $(g)$, solutions
can be described with some having an attractor behaviour.

The asymptotic behavior of $s$ near the point $(f)$ can be approximated
as $s\approx1+\exp(-\tau)$; hence, as $\tau\rightarrow\infty$ (i.e.,
$a\rightarrow0$), $s\rightarrow1$. Moreover, the time derivative
of the tachyon field is given by $\dot{\phi}\simeq1$; that is, the
tachyon field $\phi(t)$ has a linear time dependence and can be approximated
as (see figure \ref{F3b}), 
\begin{equation}
\phi(t)\ \simeq\ t+\phi_{0}.
\end{equation}

The total energy density of the collapsing system is given approximately
by the energy density of the barotropic fluid (see figure \ref{dominance},
where $\rho_{b}/\left(3H^{2}\right)\rightarrow1$ while $\rho_{\phi}/\left(3H^{2}\right)\rightarrow0$
for $\gamma>1$ ), which goes, near the singularity, as $\rho_{b}\propto1/a^{3\gamma}$.
From this relation for the fluid energy, we may write the energy density
of the system near point $(f)$, 
\begin{equation}
\rho\ \approx\ \rho_{0b}a^{-3\gamma}.\label{Tenerg-a-1}
\end{equation}
 For $\gamma<2/3$, the ratio $F/R=\frac{1}{3}r^{2}\rho_{0b}a^{2-3\gamma}$,
converges as the singularity is reached leading to the avoidance of
trapped surfaces, but since the corresponding fixed point $(f)$ turns
to be a saddle, then the resulting naked singularity is not stable.
For $2/3<\gamma<1$ the ratio $F/R$ diverges and the trapped surfaces
do form. But still the point $(f)$ is saddle and the resulting black
hole is not stable. The case $\gamma>1$ corresponds to a stable solution
for which the ratio $F/R$ goes to infinity as the collapse advances.
Then, the trapped surface formation in the collapse takes place before
the singularity formation and thus the final outcome is a black hole.

As far as $\left(g\right)$ is concerned, the time dependence of the
tachyon field can be obtained as 
\begin{equation}
\phi(t)\simeq-t+\phi_{0}.
\end{equation}
 For the $\phi>0$ branch, the tachyon field decreases from its initial
value at $t=0$ (where $\phi_{0}>0$), moving uphill the potential,
the potential will reach a maximum (but finite) value when the tachyon
field reaches its minimum and nonzero value (i.e. $\phi\rightarrow\phi_{s}$
as $t\rightarrow t_{s}\leq\phi_{0}$). In the limit case $t_{s}=\phi_{0}$,
the tachyon field vanishes and the potential diverges. I.e., $y$
and $s$ vanish, the Hubble rate increases faster than the potential
and the barotropic fluid energy density diverges, that is, the total
energy density of the system, is given by equation (\ref{Tenerg-a-1}),
and asymptotically diverges. Similar to point $(f)$ the mass function
of the system for the fixed point solution $(g)$ is given by $F/R=\frac{1}{3}r^{2}\rho_{0b}a^{2-3\gamma}$,
which diverges for $\gamma>1$. Therefore, the resulting singularity
in this case will be covered by a black hole horizon.

As far as the energy conditions are concerned for the fixed point
solutions $(f)$ and $(g)$, we find that the tachyon field satisfies
the WEC. Also the DEC remains valid case as well. On the other hand,
for the barotropic fluid, the WEC is satisfied initially and will
hold until the endstate of the collapse. The stable solution in this
case corresponds to the range $\gamma>1$ which satisfies DEC as well.

\subsubsection{Tracking solution: black hole and naked singularity formation}

\label{Tracking-solutions}

In this subsection, we will discuss a different type of solutions,
where the fluid and tachyon appear with a tracking behaviour. Let
us introduce this situation as follows. 

Interesting and physically reasonable tracking solutions can be found,
where $\dot{\phi}\rightarrow\pm1$, when we consider $\gamma\rightarrow1$,
i.e., a situation whereby the emergence of points $\left(h\right)$
and $\left(i\right)$ will be of relevance as attractors. The transition
from tachyon dominated to fluid dominated scenarios, like those described
in previous sections, is not straightforward. In this situation, the
tachyon field and barotropic fluid compete to establish the dominance
in the late stage of the collapse. In figure \ref{dominance} we have
a illustration of this kind of solutions. The mentioned dominance
seems to depend strongly on the initial ratio $\rho_{\phi}/\rho_{b}$
at an earlier stage of the collapse and also on the value of $\gamma$.
Such a dependence on the initial conditions can lead to a set of solutions
between those provided by fixed points $\left(a\right)$, $\left(b\right)$
(tachyon dominated solutions) and by points $\left(f\right)$, $\left(g\right)$
(fluid dominated solutions). From a dynamical system point of view,
this corresponds to have trajectories asymptotically approaching $x\rightarrow\pm1$
in sections where $s$ is between 0 and 1 (see figure \ref{3D-stream}).
At the end of the collapse we observe that $\frac{\rho_{\phi}}{3H^{2}}\sim\frac{\rho_{b}}{3H^{2}}<1$,
as illustrated in figure \ref{dominance}. Therefore, in this scenario,
the trajectories would convey a collapsing case in which the energy
density of the tachyon field and of the barotropic fluid are given
by $\rho_{\phi}\propto\rho_{\text{b}}\propto a^{-3\gamma}$. This
shows a tracking behavior for the collapsing system \cite{RLazkoz,Liddle}.
Moreover, the total energy density of the collapse, in terms of $a$,
reads 
\begin{equation}
\rho\ \propto\; a^{-3\gamma}.\label{energ-e}
\end{equation}

Equation (\ref{energ-e}) shows that the energy densities of the tachyon
field, the barotropic matter and hence, their total (for the collapsing
system), diverges as $a\rightarrow0$. The ratio of the total mass
function over the area radius is given by 
\begin{equation}
\frac{F}{R}\ \propto\ \frac{1}{3}r^{2}a^{2-3\gamma}.\label{massd}
\end{equation}
 Equation (\ref{massd}) subsequently implies that, for an adequate
choice of values $(\gamma,\phi_{0})$, trapped surfaces can form as
the collapse evolves and a few scenarios can be extracted. More precisely,
for the range $\gamma>\frac{2}{3}$, for both $\phi>0$ and $\phi<0$
branches, the final fate of the collapse is a black hole. For the
case in which $\gamma<\frac{2}{3}$, the ratio $F/R$ remains finite
as the collapse proceeds and an apparent horizon is delayed or fails
to form; the final state is a naked singularity (a solution for the
choice of `$-$' sign in equation (\ref{eigencle})). The tracking
solution indicates $\gamma=\frac{2}{3}$ as the threshold (illustrated
in figure \ref{2D-stream}), which distinguishes a black hole or a
naked singularity forming. 

\begin{figure*}[h]
\begin{minipage}[c]{1\textwidth}%
\centering \includegraphics[width=2.1in]{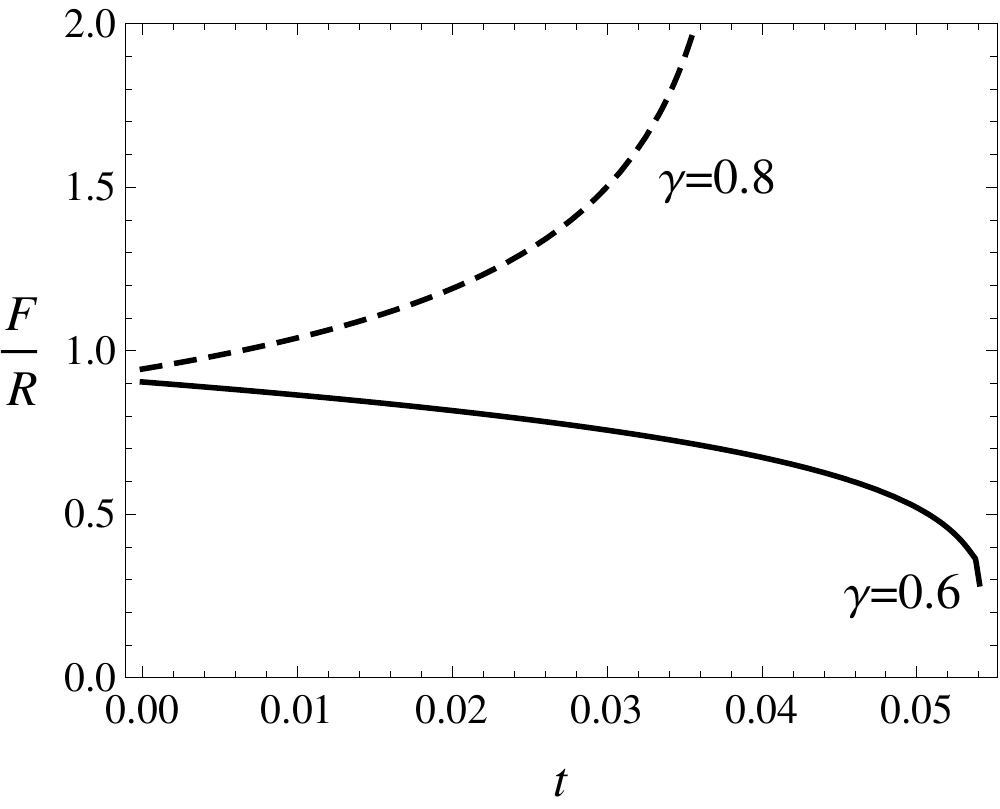}\quad{}\includegraphics[width=2.1in]{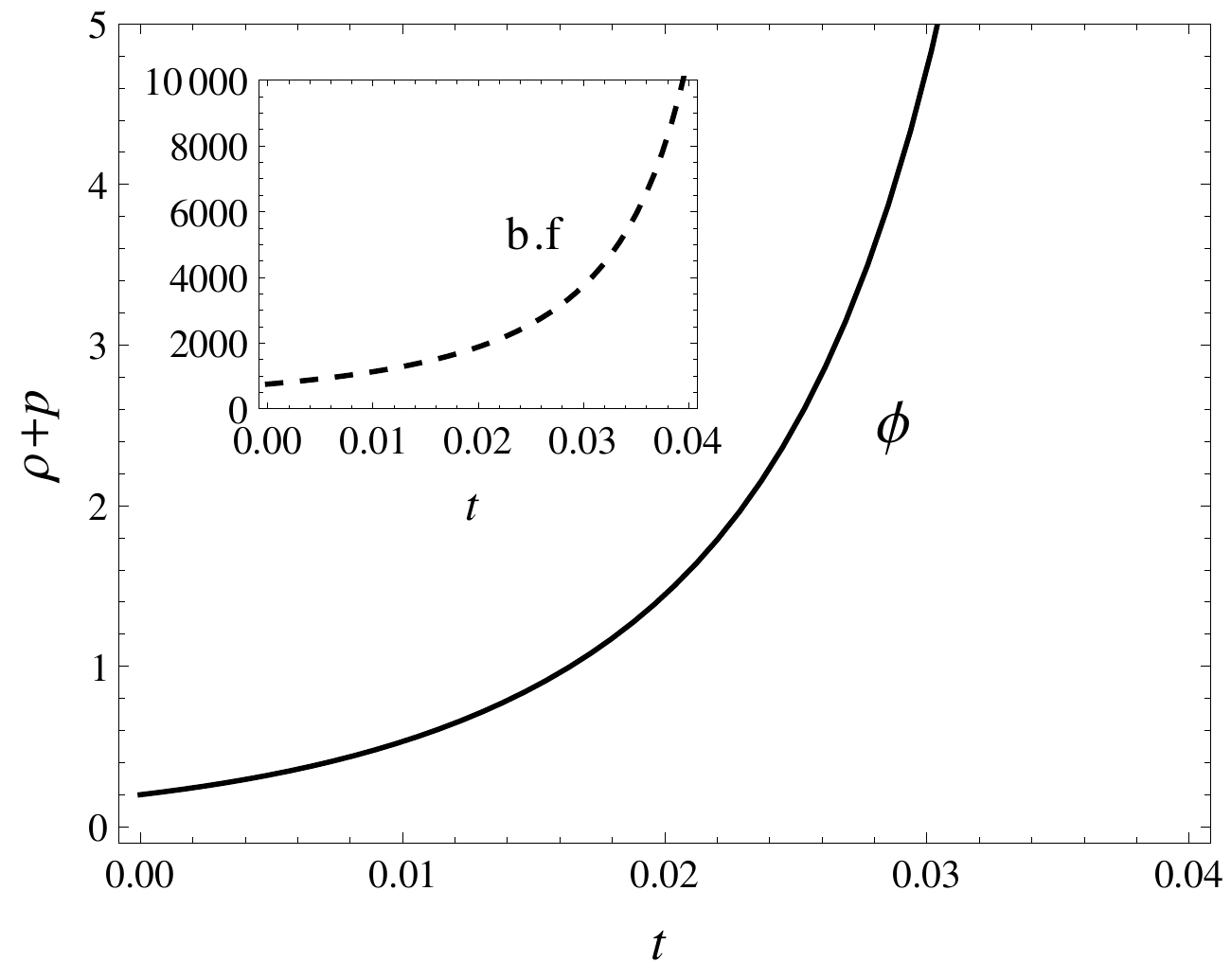}%
\end{minipage}\caption{{\footnotesize In the left plot it is shown the ratio of the total
mass function over the area radius. We considered $V_{0}=4/9$ and
$\rho_{0b}=1100$. It is shown that for $\gamma>2/3$ we have black
formation and for $\gamma<2/3$ we have naked singularity formation.
The right plot shows the weak energy condition over time for the barotropic
fluid (b.f) and for the tachyon field ($\phi$). Therein we considered
$t_{i}=0$, $a(0)=a_{0}$, $V_{0}=4/9$, $\frac{\rho_{0b}}{\rho_{0\phi}}=1100$
and $\phi_{0}=0.6$ with $\gamma=0.6$.}}

{\footnotesize \label{naked singularity} }
\end{figure*}

Therefore, under suitable conditions, we can determine whether it
is possible to have the formation of a naked singularity. In fact,
if we assume a very unbalanced initial ratio $\rho_{\phi}/\rho_{b}$
with $\rho_{0\phi}\ll\rho_{0b}$ and a barotropic fluid having $\gamma<\frac{2}{3}$,
then we can have a situation where the ratio (\ref{massd}) is converging.
The set of initial conditions described by $\rho_{0\phi}\ll\rho_{0b}$
are equivalent to consider the barotropic fluid as initially dominant.
If this specific unbalanced distribution of matter is allowed to evolve
into a regime where the tachyon dominates then the system will evolve
until $\rho_{\phi}$ becomes comparable to $\rho_{b}$. The singularity
is reached in finite time and it can happen before the tachyon can
effectively dominate. In figure \ref{naked singularity} we have a
graphical representation of the ratio $F/R$. It can be seen that
the ratio $F/R$ remains finite for $\gamma<2/3$, while the energy
density is diverging, as the collapse proceeds and apparent horizon
is delayed or fails to form till the singularity formation. As the
right plot shows, the validity of WEC is guaranteed throughout the
collapse scenario for both barotropic fluid and tachyon field. Also
the DEC is valid for the solutions that exhibit naked singularity,
i.e., those for which $\gamma<2/3$. \medskip{}

\subsubsection{Other solutions}

\label{Classic1-1new}

Point $(c)$ does not represent an attractor and hence its vicinity
will not corresponds to a collapse endstate. However, trajectories
emerging from it would proceed towards $(a)$ or $(b)$. The time
dependence of the tachyon field near $(c)$ can be obtained integrating
$\dot{\phi}=-\lambda\sqrt{\frac{y_{0}}{3}}$ as 
\begin{equation}
\phi(t)\ \simeq\ -\lambda\sqrt{\frac{y_{0}}{3}}t+\phi_{0}.
\end{equation}
 In the $\phi>0$ branch, the tachyon field starts to increase with
time from its initial value at $t=0$ (where $\phi(t=0)=\phi_{0}>0$),
proceeding uphill the potential. Then, a stage is obtained where a
maximum (but finite) value is reached, when the tachyon field goes
to its minimum and nonzero value at $t_{s}\leq\lambda\sqrt{3/y_{0}}$.
Asymptotically, $y=\frac{V}{3H^{2}}\rightarrow y_{0}$ and $s=\frac{\rho_{\text{b}}}{3H^{2}}\rightarrow0$.
Since $y_{0}>0$ is a constant, the Hubble rate at $t=t_{s}$ reads
$H=-\sqrt{V/3y_{0}}$ and the total energy density of the system is
essentially the energy density of the tachyon field and is given by
\begin{equation}
\rho\ \equiv\ 3H^{2}\ \simeq\ \frac{V(\phi)}{y_{0}}\ ,\label{cEnerg}
\end{equation}
 which is proportional to the tachyon potential. Therefore, at the
time $t_{s}<\lambda\sqrt{3/y_{0}}$, when the tachyon field reaches
its non-zero minimum, the total energy density of the system remains
finite. In the limit case, when the time $t_{s}=\lambda\sqrt{3/y_{0}}$,
the tachyon field vanishes, the tachyon potential diverges. Alternatively,
we can also write that tachyon behaves as 
\begin{equation}
\phi\ \approx\ a^{\frac{\lambda^{2}y_{0}}{2}}.
\end{equation}
 When $a\rightarrow0$, the tachyon field vanishes, and hence, the
potential and energy density of tachyon field in equation (\ref{cEnerg}),
as a function of $a$, is given by $\rho\propto V(\phi)\approx a^{-\lambda^{2}y_{0}}$
and diverges at $a=0$. The dynamics of the system in vicinity of
this fixed point is given by $x\approx-\frac{\lambda}{\sqrt{3}}\sqrt{y_{0}}$,
$y\approx y_{0}+(a_{0}/a)^{3\zeta_{2}}$ and $s\approx(a_{0}/a)^{3\zeta_{3}}$,
where $\zeta_{2}>0$.

Concerning $(d)$, the energy density of the tachyon field is approximately
given by the tachyon potential $\rho_{\phi}\simeq V(\phi)$ (with
$\dot{\phi}=0$) and thus, in its vicinity, the tachyon potential
remains constant and finite. Furthermore, $y=V/3H^{2}\simeq0$ and
$\rho_{b}/3H^{2}\simeq1$ imply that the total energy density is given
by the energy density of barotropic matter only: $\rho=3H^{2}\simeq\rho_{\text{b}}$.
This solution diverges as $a\rightarrow0$.

The fixed point $(e)$ for the range $\gamma<\gamma_{1}$ presents
solutions for the collapsing system in which $\dot{\phi}$ asymptotically
is given by $\dot{\phi}=-\sqrt{\gamma}$, for the $\phi>0$ and $\dot{\phi}=\sqrt{\gamma}$
for the $\phi<0$ branch. For the $\phi>0$ branch, the time dependence
of the tachyon field is given by 
\begin{equation}
\phi(t)=-\sqrt{\gamma}t+\phi_{0}.\label{tach-e}
\end{equation}
 The tachyon starts its evolution from the initial configuration $\phi_{0}>0$,
and decreases uphill the potential. Then, at the time $t=t_{s}=\frac{\phi_{0}}{\sqrt{\gamma}}$
the tachyon field vanishes, the potential diverging%
\footnote{For the $\phi<0$ branch, the time dependence of the tachyon field
is given by $\phi(t)=\sqrt{\gamma}t+\phi_{0}$. Therefore the tachyon
field starts its evolution from the initial configuration at $\phi_{0}<0$,
decreasing in time, going uphill the potential.%
}. 

A thorough numerical analysis of the trajectories in the vicinity
of point $\left(e\right)$ has shown that it behaves as an unstable
node. We can observe in the center plot of figure \ref{2D-stream}
an illustration of the general behavior near this point, namely having
trajectories diverging from it.

\subsection{Exterior geometry}

\label{Classic3}

In order to complete the full space-time geometry for the herein collapsing
model, we need to match the homogeneous interior space-time to a suitable
(inhomogeneous) exterior geometry. For a perfect fluid gravitational
collapse set up, with equation of state $p=\left(\gamma-1\right)\rho$,
the pressure does not necessarily vanish at the boundary. E.g., matching
the internal geometry filled with matter (and radiation) to a boundary
layer (which is crossed with the radiation), which could in turn be
matched to an exterior geometry, not completely empty (e.g., filled
by radiation). More concretely, matching the interior with a generalized
exterior Vaidya space-time across the boundary given by $r=r_{b}$
\cite{Joshi}.

Similarly to \cite{Hussain}, we proceed by considering the interior
metric which describes the collapsing cloud as 
\begin{align}
ds_{-}^{2}=-dt^{2}+a^{2}(t)(dr^{2}+r^{2}d\Omega^{2})\label{int}
\end{align}
 and the exterior one in advanced null coordinates $(v,R)$ given
by 
\begin{align}
ds_{+}^{2}=-f(v,R)dv^{2}+2dvdR+R^{2}d\Omega^{2}.\label{out}
\end{align}
 In these coordinates the outermost boundary of trapped surfaces is
simply given by the relation $f(v,R)=0$. It should be noticed that
in general the formation of the singularity at $a=0$ is independent
of matching the interior to the exterior space-time. In order to find
a suitable exterior metric function $f(v,R)$ we resort to an Hamiltonian
perspective of the model. The total Hamiltonian constraint is given
by \cite{Sen} 
\begin{align}
{\cal {H}}(a,\pi_{a};\pi_{\phi}) & ={\cal {H}}_{a}+{\cal {H}}_{\phi}+{\cal {H}}_{\text{b}}\notag\\
 & =\frac{\pi_{a}^{2}}{12a}\ +\ a^{3}\sqrt{V^{2}+{a}^{-6}\pi_{\phi}^{2}}\ +\ \rho_{0b}a^{-3\left(\gamma-1\right)},\label{Hamiltonian-cl}
\end{align}
 where $\pi_{a}$ and $\pi_{\phi}$ are the conjugate momentums for
the scale factor $a$ and for the tachyon field $\phi$, respectively.
Furthermore, the Hamilton equations for parameters $a$, $\pi_{a}$
and $\pi_{\phi}$ can be obtained by using the equation (\ref{Hamiltonian-cl})
as follows: 
\begin{align}
\dot{a} & =\frac{\partial{\cal {H}}}{\partial\pi_{a}}=\frac{\pi_{a}}{6a},\ \ \ \ \ \ \dot{\phi}=\frac{\partial{\cal H}}{\partial\pi_{\phi}}=\frac{a^{-3}\pi_{\phi}}{\sqrt{V^{2}+a^{-6}\pi_{\phi}^{2}}}\ ,\label{eqs-1}\\
\dot{\pi}_{a} & =-\frac{\partial{\cal {H}}}{\partial a}=\frac{\pi_{a}^{2}}{12a^{2}}+3\left(\gamma-1\right)\rho_{0\text{b}}a^{-(1+3\left(\gamma-1\right))}-\frac{3a^{5}V^{2}}{\sqrt{a^{6}V^{2}+\pi_{\phi}^{2}}}\ ,\label{eqs-2}\\
\dot{\pi}_{\phi} & =-\frac{\partial{\cal {H}}}{\partial\phi}=-\frac{a^{3}VV_{,\phi}}{\sqrt{a^{6}V^{2}+\pi_{\phi}^{2}}}\ ,\label{eqs-3}
\end{align}
 On the other hand, since the Hamiltonian constraint, ${\cal {H}}=0$,
must be held across the boundary $\Sigma$, equation (\ref{Hamiltonian-cl})
reduces to 
\begin{equation}
\pi_{\phi}^{2}=\frac{\pi_{a}^{4}}{144a^{2}}+\rho_{0\text{b}}^{2}a^{-6\left(\gamma-1\right)}+\rho_{0\text{b}}\frac{\pi_{a}^{2}a^{-3\left(\gamma-1\right)}}{6a}-a^{6}V^{2},\label{pphi}
\end{equation}
 whereby, substituting for $\pi_{a}$ and $\pi_{\phi}$ from equation
(\ref{eqs-1})-(\ref{eqs-3}), we get 
\begin{equation}
9a^{2}\dot{a}^{4}+\rho_{0b}^{2}a^{-6\left(\gamma-1\right)}+6\rho_{0b}\dot{a}^{2}a^{2-3\gamma}=\frac{a^{6}V^{2}}{1-\dot{\phi}^{2}}\ .\label{fex}
\end{equation}
 Furthermore, it is required for the junction condition $r_{b}a(t)=R(t)$
and the equation of motion for scale factor $r_{b}\dot{a}=\dot{R}$
to be satisfied at the boundary of two regions. By substituting these
conditions in the Hamiltonian constraint equation, ${\cal H}(v,R)=0$,
given by (\ref{fex}), we get 
\begin{align}
(1-f)\left[9(1-f)+6\rho_{0\text{b}}\left(\frac{R}{r_{b}}\right)^{4-3\gamma}\right]=\left(\frac{R}{r_{b}}\right)^{4}\left[\frac{V^{2}}{1-\dot{\phi}^{2}}-\rho_{0\text{b}}^{2}\left(\frac{R}{r_{b}}\right)^{-6\gamma}\right],\label{AREA}
\end{align}
 Notice that, we have substituted the four-velocity of the boundary
being seen from the exterior by $\dot{R}=-\sqrt{1-f}$ (cf. \cite{Hussain}).
Then, by solving equation (\ref{AREA}) for $f$, the boundary function
is obtained simply as 
\begin{equation}
f(R)=1-\frac{2}{3}\frac{V}{\sqrt{1-\dot{\phi}^{2}}}\left(\frac{R}{r_{b}}\right)^{2}+\frac{1}{3}\rho_{0\text{b}}\left(\frac{r_{b}}{R}\right)^{1+3\left(\gamma-1\right)}.\label{exteriormf}
\end{equation}

Let us now study the behavior of the boundary function for the stable
fixed point solutions we obtained in the previous subsections. We
note that the above expression is valid when both the tachyon field
and barotropic fluid are present, thus in the regimes where the tachyon
field is dominant the contribution of the fluid to Hamiltonian constraint
is set aside. For the dust-like solutions, described by points $\left(a\right)$
and $\left(b\right)$, the energy density of tachyon field at the
boundary $r_{b}$, included in second term of equation (\ref{exteriormf}),
is given by $\rho_{\phi}\approx\rho_{o}r_{b}^{3}/R^{3}$. So, the
boundary function $f$ in equation (\ref{exteriormf}) reduces to:
\begin{equation}
f(R)=1-\frac{M}{R},\qquad\text{where}\qquad M\equiv\frac{4\pi}{3}r_{b}^{3}\tilde{\rho}_{o},\label{exteriormf-dust}
\end{equation}
 and $\tilde{\rho}_{o}\equiv\rho_{o}/2\pi r_{b}^{2}=\text{const}$.
Equation (\ref{exteriormf-dust}) for $f(R)$ constrains the exterior
space-time to have a Schwarzschild metric in advanced null coordinates,
providing an interpretation of the collapsing system as a dust ball
with the radius $r_{b}$ and the density $\tilde{\rho}_{o}$. On the
other hand, for the fixed point solution $(e)$, by using the equation
(\ref{energ-e}) in (\ref{exteriormf}), we have 
\begin{equation}
f(R)=1-\frac{\tilde{M}}{R^{3\gamma-2}},\qquad\text{where}\qquad\tilde{M}\equiv\frac{2}{3}\left(\rho_{0}-\frac{2s_{0}+1}{s_{0}}\rho_{0\text{b}}\right)r_{b}^{3\gamma-2}=\text{const},\label{exteriormf-e}
\end{equation}
 where $3\gamma-2>0$ (notice that, this corresponds to the case in
which $\gamma=\gamma_{1}>2/3$ in previous subsection).

In order to get a possible class of dynamical exterior solutions,
we proceed by considering the following metric at the boundary \cite{Hussain}
\begin{equation}
ds^{2}=-\left(1-\frac{2M(R,v)}{R}\right)dv^{2}+2dvdR+r^{2}d\Omega^{2},\label{bm}
\end{equation}
 where 
\begin{equation}
M(R,v)=m(v)-\frac{g(v)}{2(2\gamma-3)R^{2\gamma-3}}\ ,\label{mass}
\end{equation}
 is the total mass within the collapsing cloud. Matching the for exterior
metric function gives 
\begin{align}
 & 1-\frac{2m(v)}{R}+\frac{g(v)}{(2\gamma-3)R^{2\gamma-2}}=1-\frac{\tilde{M}}{R^{3\gamma-2}}\ ,\notag\\
 & \frac{2\dot{m}(v)}{R}=\frac{\dot{g}(v)}{(2\gamma-3)R^{2\gamma-2}}\ ,\label{matching}
\end{align}
 in which, the second part stands for matching for the extrinsic curvature.
Differentiation of the first expression in (\ref{matching}) with
respect to time and using the second one, we get 
\begin{equation}
\frac{2m(v)}{R^{2}}-\frac{2(\gamma-1)g(v)}{(2\gamma-3)R^{2\gamma-1}}=\frac{\tilde{M}(3\gamma-2)}{3R^{3\gamma-1}}\ .\label{1}
\end{equation}
 Multiplying the first expression in equation (\ref{matching}) by
$R^{-1}$ and after adding the result with the above equation, we
get 
\begin{equation}
g(v)=-\frac{\tilde{M}(\gamma-1)}{R(v)^{\gamma}}\ .\label{g}
\end{equation}
 Now, by substituting for $g(v)$ into equation (\ref{matching}),
we obtain 
\begin{equation}
m(v)=\frac{\tilde{M}\gamma}{6(3-2\gamma)R(v)^{3\gamma-3}}\ .\label{m}
\end{equation}
 Then, as seen from equation (\ref{massd}) for $\gamma<2/3$, where
the trapping of light has failed to occur, the exterior geometry is
dynamical in contrast to the tachyon dominated regime in which a space-like
singularity forms with a static exterior space-time. Finally, for
fluid dominated solutions, depending on the value of $\gamma$, both
naked singularities and black holes may form, the mass being different
to those of tracking solutions and the exterior geometry being static
or dynamical, respectively.

\section{Conclusions, Discussion and Outlook}

\label{discussion}

In this paper we considered a particular setting among the spherically
symmetric class of models for gravitational collapse, with a tachyon
field and a barotropic fluid as matter content. We restricted ourselves
to the marginally bound case (cf. \cite{2,4,1} for a description
and details). The tachyon potential was assumed to be of an inverse
square form. Our objective was to establish (i) which final states
would occur (namely, a black hole or a naked singularity), (ii) how
each matter component will compete (the fluid being conventional,
whereas the tachyon bringing some workable but intrinsic non-standard
effects from string theory) and, (iii) which will eventually be the
determinant component at the end of the collapse process. More precisely,
in our opinion it is of interest to assert, if and how, at later stages,
effects induced by the tachyon (a scalar field (among others) found
in string theory context), could allow interesting features to be
eventually discussed (i.e., what asymptotic behaviour emerges). As
far as the tachyon field is concerned, from eq. (\ref{field}) we
can have that with $-1<\dot{\phi}<1$ and $H<0$, $\dot{\phi}$ terms
act like an anti-friction contribution at a collapsing phase, within
an uphill evolution for the $\phi^{-2}$ potential for tachyon field
(i.e., when $\dot{\phi}<0$). Moreover, other terms (i.e. the term
including $\frac{V_{,\phi}}{V}$) would have an anti-friction effect
as well.

Determining therefore the outcome of the gravitational collapse in
our system%
\footnote{The case of a standard scalar field was investigated in \cite{1,4}%
}, i.e., whether, e.g., a black hole or naked singularity would form,
was not a straightforward assessment. We considered an analytical
description by means of a phase space analysis \cite{Khalil,RLazkoz},
discussing several asymptotic behaviours; these were also subject
to a careful study involving a numerical investigation, which added
a clearer description of the possible dynamical evolutions. Within
this setting, for a spatially homogeneous interior space-time, we
found a situation where the tachyon was dominant, with a black hole
forming%
\footnote{It is worthwhile to compare the result obtained herein for a homogeneous
(tachyonic) collapsing matter field with the gravitational collapse
of a k-essence scalar field with non-standard kinematic terms in \cite{AGS},
where the scalar field has a dependence to the radius $r$ i.e., $\phi=\phi(r)$.
In both models, the collapsing systems lead to the black hole formation.%
}. A cosmological framework involving only a FRW geometry, with a tachyon
field and a barotropic fluid, was investigated in \cite{RLazkoz},
focusing on the late time (dark energy like) stages. It is interesting
to note that while for a tachyon dominated regime, an inflationary-like
(i.e., accelerated expansion) scenario leads to violation of the strong
energy condition (SEC)%
\footnote{Satisfying the SEC for the tachyon field demands that $\rho_{\phi}+3p_{\phi}\geq0$.
Thus, the SEC holds if $\dot{\phi}^{2}>2/3$.%
} \cite{RLazkoz}, in our collapse scenario we have instead a corresponding
asymptotic stage, where tachyon dominance leads to black hole formation,
satisfying the SEC at the final state of the collapse (cf. figure
(\ref{F3b})). The same behaviour is observed in the fluid dominated
regime, where for $\gamma>1$ the SEC is satisfied by the fluid, with
a black hole formation. However, for the tracking solutions, where
a naked singularity forms, i.e., $\rho_{0b}>>\rho_{0\phi}$ and $\gamma<2/3$,
the SEC is violated.

Moreover, in further comparison with the set of fixed points found
in \cite{RLazkoz} for an expanding (accelerating) universe ($H>0$),
we have found analytically two additional critical points ($f$ and
$g$) for the collapse process which correspond to a barotropic dominated
collapsing regime which ends in a black hole for $\gamma>1$, respecting
the WEC and SEC. Indeed, for the late-time acceleration of the universe,
filled with a tachyon field and a fluid, the phase space analysis
in \cite{RLazkoz} predicted two class of solutions: a tachyon dominated
solution; and a tracking solution \cite{RLazkoz}. Rather differently,
corresponding solutions herein our paper turn to be unstable towards
the singularity in the collapse process. Nevertheless, two other solutions
are also provided in our collapsing scenario: a tachyon dominated
solution where $\gamma$ satisfies the range $\gamma<1$, initially;
and a barotropic dominated solution for which the barotropic parameter
holds the range $\gamma>1$. All these solutions predict a black hole
formation as collapse end state.

Further regarding the tracking solutions (of a cosmological nature)
indicated in \cite{RLazkoz}. In our collapsing system, a different
and rather interesting set of states is found, within the context
of tracking behavior for the barotropic fluid plus tachyon field content.
Being more concrete, these solutions have that a black hole or naked
singularity forms. In particular, in this situation, we found that
it is possible to define by a numerical appraisal in particular, the
threshold $\gamma=2/3$, separating black hole and naked singularity
solutions for the gravitational collapse. We have also discussed the
specific conditions leading to the formation of a naked singularity.
Therefore, we concluded, that in our model, if the collapse starts
with an unbalanced distribution of the matter content favoring the
barotropic fluid, i.e. $\rho_{\phi}\ll\rho_{b}$, then towards the
final stage of the collapse they evolve until $\rho_{\phi}\approx\rho_{b}$,
when a naked singularity forms. However, since the NEC is satisfied,
the strong curvature condition along the null geodesics can be preserved
and the singularity can be strong in the sense of \cite{Tipler 1}.

We think it is fair to indicate that we employed a $\phi^{-2}$ potential
for the tachyon, whereas for $\phi\rightarrow0$, the tachyon should
not induce a divergent behavior as far as string theory advises \cite{Feinstein,DynTach,31a,31b,31c}.
In fact, an exponential-like potential for the tachyon could bring
a richer set of possible outcomes \cite{dbi,Feinstein,AsenH,ASen2},
including a better behaved and possibly a more realistic evolution
when dealing with $\phi\rightarrow0$.

Finally, let us add that it will be of interest to investigate (i)
other scenarios for the geometry of the interior space-time region,
(ii) specific couplings between the tachyon and the fluid, within
e.g., a chamaleonic scenario for gravitational collapse and black
hole production (broadening the scope in \cite{Folomeev:2012sz,Folomeev:2011aa,Folomeev:2011uj}),
(iii) adding either axionic, dilatonic or other terms (e.g., curvature
invariants) that could be considered from a string setup, but at the
price of making the framework severely less workable and (iv) whether
explicit quantum effects can alter the outcomes presented in this
paper. To this purpose latter, we plan to use ingredients brought
from loop quantum gravity (cf. ref. \cite{THMM-2}).

\section{Acknowledgments}

The authors would like to thank P. Joshi, A. Khaleghi, C. Kiefer,
F. C. Mena, A. Vikman, and J. Ward for making useful suggestions and
comments. They are also grateful to the referee for the useful comments
and suggestions on the issue of energy conditions. YT is supported
by the Portuguese Agency Fundação para a Ciência e Tecnologia through
SFRH/BD/43709/2008. This research work was also supported by the grant
CERN/FP/109351/2009 and CERN/FP/116373/2010.

\end{document}